\documentclass[a4paper,11pt]{article}
\usepackage{jheppub} % for details on the use of the package, please see the JINST-author-manual
\usepackage{dsfont}
\usepackage{tikz}
\usetikzlibrary{braids,knots}
\usepackage{mathtools}
\usepackage{caption}
\title{\boldmath Hadamard products and BPS networks}

\author{Mohamed Elmi}
\affiliation{NHETC, Department of Physics and Astronomy\\
Rutgers University\\
136 Frelinghuysen Road\\
Piscataway, NJ 08855\\
USA
}

\emailAdd{elmi@physics.rutgers.edu}

\abstract{We study examples of fourth-order Picard-Fuchs operators that are Hadamard products of two second-order Picard-Fuchs operators. Each second-order Picard-Fuchs operator is associated with a family of elliptic curves, and the Hadamard product computes period integrals on the fibred product of the two elliptic surfaces. We construct $3$-cycles on this geometry as the union of $2$-cycles in the fibre over contours on the base. We then use the special Lagrangian condition to constrain the contours on the base. This leads to a construction that is reminiscent of spectral networks and exponential networks that have previously appeared in string theory literature.}

\begin{document}
\maketitle
\flushbottom
\section{Introduction}

Almkvist, Enckevort, van Straten and Zudilin have compiled a list of fourth order, linear differential operators of {\em Calabi-Yau type} that we will refer to as the {\em AESZ list}. \cite{2005math......7430A} By definition, these operators satisfy a number of conditions that make them candidate Picard-Fuchs operators for a family of Calabi-Yau threefolds with Hodge number $h^{2,1}=1$. For example, each operator should have a point of maximal unipotent monodromy at $\varphi=0$, there should exist a power series solution at $\varphi=0$ with integer coefficients, etc.

There are many operators in the AESZ list that are the {\em Hadamard product} of two second order Picard-Fuchs equations, each associated with a family of elliptic curves. They are defined as follows. Let $\mathcal{L}^{(1)}$ and $\mathcal{L}^{(2)}$ be Picard-Fuchs operators in the variable $s$ such that
\begin{equation}
\mathcal{L}^{(i)}\left(\sum_{n=0}^{\infty}a_n^{(i)}s^n\right)=0.
\end{equation}
We say that a non-trivial differential operator $\mathcal{L}$ is a {\em Hadamard product} of $\mathcal{L}^{(1)}$ and $\mathcal{L}^{(2)}$ if it satisfies
\begin{equation}
\label{eq: Hadamard product definition}
    \mathcal{L}\left(\sum_{n=0}^{\infty} a_n^{(1)}a_n^{(2)} \varphi^n\right)=0.
\end{equation}
If $\mathcal{L}^{(1)}$ and $\mathcal{L}^{(2)}$ are of finite order with polynomial coefficients, there exists an operator $\mathcal{L}$ that satisfies \eqref{eq: Hadamard product definition}.\cite{{andre1989g}} However, this operator is not unique, and we know of no general algorithm for computing it. Moreover, there is no guarantee that the Hadamard product of two second-order Calabi-Yau operators will be a fourth-order Calabi-Yau operator. In practice, we simply make the ansatz
\begin{equation}
\mathcal{L}=S_4(\varphi)\theta^4 + S_3(\varphi)\theta^3 + S_2(\varphi)\theta^2 + S_1(\varphi)\theta^1+S_0(\varphi),
\end{equation}
where $\theta=\varphi\frac{d}{d\varphi}$ and $S_i(\varphi)$ is a polynomial in $\varphi$. We then insist that it satisfy \eqref{eq: Hadamard product definition} and solve for the coefficients in the polynomials $S_i(\varphi)$. In all the examples we study here, there will be a differential operator $\mathcal{L}$, unique up to some overall scalar multiple, of minimal degree in the polynomials $S_i(\varphi)$ that satisfies \eqref{eq: Hadamard product definition}. We will refer to this operator as the Hadamard product of $\mathcal{L}^{(1)}$ and $\mathcal{L}^{(2)}$.

Geometrically, the Hadamard products compute the periods of a holomorphic $3$-form on the fibred product of the two families of elliptic curves over a common $\mathbb{CP}^1$ with punctures, where the structure of the fibration and the location of the punctures depend on a complex parameter $\varphi$ (see, for example, the thesis of Samol \cite{DissertationSamol}). We may construct $3$-cycles on the fibred product as the union of $S^1\times S^1$ over some contour on the base $\mathbb{CP}^1$. Here, each $S^1$ must lie in one of the elliptic fibres and both the contour on the base and the homology classes in the fibres must be carefully chosen to ensure that the $2$-boundary vanishes. We may then integrate the holomorphic $3$-form over this $3$-cycle. The resulting integral will be a function of $\varphi$ and will be annihilated by the Hadamard product. When the Hadamard product is a Picard-Fuchs equation of a family of compact Calabi-Yau threefolds, the same integrals may be identified with periods of the holomorphic $3$-form on a compact Calabi-Yau threefold. This gives us a precise map between $3$-cycles on the fibred product of elliptic curves and $3$-cycles on the Calabi-Yau threefold. As examples, we will use the Picard-Fuchs operators AESZ 3 and AESZ 101, which are both Hadamard products. 

In Sections~\ref{section: AESZ 3} and~\ref{section: AESZ 101}, we compute solutions of AESZ 3, solutions of AESZ 101 and solutions of the associated second order Picard-Fuchs equations in appropriate bases. We also explain, for each example, the precise relationship between the family of smooth Calabi-Yau threefolds and a fibred product of families of elliptic curves. We then turn to the subject of special Lagrangian submanifolds.

A D3 brane on a Calabi-Yau threefold at zero string coupling is described by a special Lagrangian submanifold with a flat $U(1)$ connection.\cite{Becker:1995kb} The special Lagrangian condition is a constraint involving the K\"{a}hler form and the holomorphic $3$-form of the Calabi-Yau threefold. We do not have much control over the K\"{a}hler form, but we do have a great deal of control over the holomorphic $3$-form, so we apply this second condition to the $3$-cycles on the fibred product of elliptic surfaces. This leads to a construction that is very similar to the spectral networks and exponential networks that have previously appeared in string theory literature.\cite{Klemm:1996bj,Gaiotto:2012rg,Eager:2016yxd} 

The networks that appear in this paper are defined as follows. Let $s$ be an in-homogeneous coordinate on $\mathbb{CP}^1$ and let $\gamma_1$ and $\gamma_2$ be homology $1$-cycles in the first and second elliptic fibres, respectively. We integrate the holomorphic $1$-forms of the first and second elliptic fibres at $s$ over $\gamma_1$ and $\gamma_2$, respectively. We will see in following sections that this will result in two periods $\omega_{\gamma_1}(s)$ and $\widetilde{\omega}_{
\gamma_2}\left(\frac{\varphi}{s}\right)$, respectively. Note the dependence on $\varphi$ and the coordinate $s$. We construct a $3$-cycle by taking the union of products of circles over contours on the base $\mathbb{CP}^1$. If the circles in the fibre are in the homology classes $\gamma_1$ and $\gamma_2$, respectively, the special Lagrangian condition requires that the contour on the base satisfy
\begin{equation}
    \omega_{\gamma_1}(s)\widetilde{\omega}_{\gamma_2}\left(\frac{\varphi}{s}\right)\frac{1}{s}\frac{ds}{dl}=e^{i\theta}
\end{equation}
for some parameterisation $s=s(l)$ and an angle $\theta$.

We have chosen to focus on Hadamard products and the associated fibred products of elliptic surfaces for their simplicity, but we expect that the ideas in this paper will generalise. We also point out that much of our analysis only depends on the structure of the associated Picard-Fuchs operators. Knowledge of the the triple intersection number, second Chern-class and Euler characteristic of the mirror Calabi-Yau manifold is used to fix a suitable basis of solutions at the point of maximal unipotent monodromy $\varphi=0$. However, even this topological data is strongly constrained by the requirement of integral symplectic monodromy and, given a fourth order Picard-Fuchs equation, we may compute the finitely many choices of topological data consistent with integral symplectic monodromy.\cite{2004math.....12539V}

\section{AESZ 3}
\label{section: AESZ 3}

\subsection{Family of Calabi-Yau threefolds}

Solutions of the Picard-Fuchs equation AESZ 3 determine periods of the holomorphic $3$-form on the mirror of the intersection of four quadrics in $\mathbb{CP}^7$ (the precise differential equation may be found in the next subsection). AESZ 3 is a Hadamard product of the Picard-Fuchs equations of two families of elliptic curves, so one might ask - how is the mirror of the intersection of four quadrics related to the fibred product of two elliptic surfaces? An answer may be found in \cite{Bonisch:2022mgw}. We summarise their computation here, with slight modifications.

The mirror of the intersection of four quadrics in $\mathbb{CP}^7$ is constructed via a Green-Plesser like construction.\cite{Greene:1990ud} One starts with an especially symmetric family, defined as the intersections of the four quadrics
\begin{equation}
    W_\psi: P_j=x_j^2+y_j^2-2\psi x_{j+1}y_{j+1}=0,
\end{equation}
where $j\in\{ 0,1,2,3\}$ and $\psi$ is a complex parameter of the family. The fibre at a given $\psi$ is invariant under the group
\begin{equation}
\Gamma=\{\xi=(\xi_j)_{j=0,...,3}\in (\mu_4)^4\}/\mu_4^{\text{diag}}
\end{equation}
where $\mu_4$ is the group generated by the primitive fourth roots of unity. $\Gamma$ is isomorphic to $(\mathbb{Z}/4\mathbb{Z})^3$ and acts on the coordinates of $\mathbb{CP}^7$ as
\begin{equation}
    (x_j,y_j)\mapsto(\xi_j x_j,\xi_{j-1}^2\xi_j^{-1} y_j).
\end{equation}
Away from $\varphi=\frac{1}{(2\psi)^8}\in\{0,\frac{1}{2^8},\infty\}$, the quotient of $W_\psi$ by $\Gamma$ may be resolved to a smooth Calabi-Yau threefold with Hodge numbers $h^{1,1}=65$ and $h^{2,1}=1$. Now consider the hypersurface in $(\mathbb{CP}^1)^4$ defined by
\begin{equation}
    \label{eq: equation for Wtilde}
    \widetilde{W}_\phi:\prod_{i=1}^4\left(Y_i-\frac{1}{Y_i}\right)=\phi.
\end{equation}
There is a degree $8$ map from $W_\psi/\Gamma$ to $W_\phi$, given by
\begin{equation}
    \label{eq: map from mirror of four quadrics in P^7 to fibred product of elliptic curves}
    iY_j=\frac{x_j-y_j}{x_j+y_j}
\end{equation}
where $j\in\{ 0,1,2,3\}$ and $\varphi=\left(\frac{\phi}{2^8}\right)^2=(2\psi)^{-8}$. $\widetilde{W}_\phi$ is a fibred product of two families of elliptic curves. To see this, note that
\begin{equation}
    \label{eq: equation for the family of elliptic curves E_s}
    \mathcal{E}_s:\left(Y_1-\frac{1}{Y_1}\right)\left(Y_2-\frac{1}{Y_2}\right)=s
\end{equation}
defines a family of elliptic curves with parameter $s$. Thus, if $Y_1$ and $Y_2$ are constrained to lie on the elliptic curve $\mathcal{E}_s$, then $Y_3$ and $Y_4$ are constrained by \eqref{eq: equation for Wtilde} to lie on the elliptic curve $\mathcal{E}_{\frac{\phi}{s}}$.

Finally, note that the families of elliptic curves $\mathcal{E}_s$ and threefolds $\widetilde{W}_\phi$ both admit the action of an involution. $\mathcal{E}_s$ is isomorphic to the curve $\mathcal{E}_{\frac{2^4}{s}}$ via the map
\begin{equation}
    \label{eq: symmetry of E_s}
    Y_i\mapsto \frac{Y_i-1}{Y_i+1}.
\end{equation}
By applying this map to each factor in \eqref{eq: equation for Wtilde}, we see that $\widetilde{W}_\phi$ is equivalent to $\widetilde{W}_{\frac{2^8}{\phi}}$.

\subsection{Solutions of AESZ 3 and a double cover}
AESZ 3 is the operator
\begin{equation}
    \label{eq: AESZ 3}
    \mathcal{L}_{\text{AESZ3}}=(1-2^8\varphi)\theta_{\varphi}^4-2^9\varphi\theta_{\varphi}^3-(2^7\cdot 3)\varphi\theta_{\varphi}^2-2^7\varphi\theta_{\varphi}-2^4\varphi,
\end{equation}
where $\theta_{\varphi}=\varphi\frac{d}{d\varphi}$. It defines a hypergeometric differential equation with the Riemann symbol
\begin{equation}
    \mathcal{P}\left\{\begin{array}{c c c} 0 & 1/2^8 & \infty\\
    \hline
    0 & 0 & 1/2\\
    0 & 1 & 1/2\\
    0 & 1 & 1/2\\
    0 & 2 & 1/2
    \end{array}~; \varphi\right\}~.
\end{equation}
A basis of solutions of around the point of maximum unipotent monodromy (MUM) at $\varphi=0$ is given by
\begin{equation}
    \label{eq: Frobenius basis at MUM point of fourth order PF equation}
    \varpi(\varphi)= 
    \left(\begin{array}{l}
    F_0(\varphi) \\ 
    F_0(\varphi) \log(\varphi) +F_1(\varphi)\\ 
    F_0(\varphi) \log^2(\varphi) +2F_1(\varphi)\log(\varphi)+F_2(\varphi)\\
    F_0(\varphi)\log^3(\varphi)+3F_1(\varphi)\log^2(\varphi)+3F_2(\varphi)\log(\varphi)+F_3(\varphi)
    \end{array}\right)
\end{equation}
where
\begin{align}
\begin{split}
F_0(\varphi)&=\sum_{n=0}^\infty\binom{2n}{n}^4 \varphi^n \\
&= 1+16\varphi+1296\varphi^2+160000\varphi^3+24010000\varphi^4+4032758016 \varphi ^5+\ldots\\
F_1(\varphi)&=64\varphi+6048\varphi^2+\frac{2368000}{3}\varphi^3+\frac{365638000}{3}\varphi^4+\frac{104147576064}{5}\varphi^5+\ldots\\
F_2(\varphi)&=64\varphi+11664\varphi^2+\frac{16364800}{9}\varphi^3+\frac{2748199300}{9}\varphi^4+\frac{1370195072064}{25}\varphi^5\ldots \\
F_3(\varphi)&=-384\varphi-25776\varphi^2-\frac{20668160}{9}\varphi^3-\frac{2221690310}{9}\varphi^4\ldots~.
\end{split}
\end{align}
We refer to this basis of solutions as the {\em Frobenius basis}. Let $\Omega$ be the holomorphic $3$-form on the mirror of the intersection of four quadrics in $\mathbb{CP}^7$ and define 
\begin{equation}
\Pi(\varphi)=\left(\int_{A_I}\Omega(\varphi),\int_{B^I}\Omega(\varphi)\right)^T,
\end{equation}
where $A_I, B^J\in H_3(X,\mathbb{Z})$ are generators that satisfy $A_I\cap B^J=-B^J\cap A_I=\delta_I^J$ with all other intersections vanishing. Such a homology basis is implicitly determined by the pre-potential of the complexified K\"{a}hler moduli space of a mirror Calabi-Yau manifold.\cite{Candelas:1990rm} The change of basis from Frobenius basis to an integral symplectic basis is given by
\begin{equation}
    \label{eq: Pi = rho*varpi}
    \Pi(\varphi)=\rho\varpi(\varphi),
\end{equation}
where
\begin{equation}
\label{eq: rho matrix for AESZ 3}
    \rho= (2\pi i)^3\begin{pmatrix} 
    \frac{\zeta(3)}{(2\pi i)^3}\chi & \frac{c_2\cdot H}{24} & 0 & \frac{H^3}{3!}\\
    \frac{c_2\cdot H}{24} & \frac{\sigma}{2} & -\frac{H^3}{2!} & 0\\
    1 & 0 & 0 & 0\\
    0 & 1 & 0 & 0
    \end{pmatrix}\begin{pmatrix} 1 & 0 & 0 & 0\\ 0 & (2\pi i)^{-1} & 0 & 0\\ 0 & 0 & (2\pi i)^{-2} & 0\\ 0 & 0 & 0 & (2\pi i)^{-3}\end{pmatrix}
\end{equation}
and 
\begin{equation}
    \sigma = \begin{cases} ~0& \text{if $H^3$ is even}\\ ~1& \text{if $H^3$ is odd}.\end{cases}
\end{equation} See, for example, \cite{Braun:2015jdy}. In our example, the topological data of the of the mirror manifold are given by $H^3=16$, $c_2\cdot H=64$ and $\chi = -128$. We define the monodromy matrices by choosing a base point in the upper half $\varphi$-plane and analytically continuing $\Pi$ in an anti-clockwise fashion around $\varphi_*$ so that $\Pi\mapsto M_{\varphi_*}\Pi$. These monodromy matrices are given by\footnote{AESZ 3 is hypergeometric, so the monodromy matrices may be computed via the Mellin-Barnes integral, as in \cite{Candelas:1990rm}. In more general examples, as in the case of AESZ 101, other methods are needed. One may simply evaluate $\Pi$ within the radius of convergence around zero, numerically integrate the solutions around a closed loop and compare with the original values. Since the monodromy matrices are integral, only a single decimal place of precision is required to fix the monodromy matrices exactly. See \cite{Braun:2015jdy} for an example. }
\begin{align}
\begin{split}
    M_{\varphi=0} &= \begin{pmatrix}
        1 & -1 & 8 & 8\\
        0 & 1 & -8 & -16\\
        0 & 0 & 1 & 0\\
        0 & 0 & 1 & 1
    \end{pmatrix},~~~~
    M_{\varphi=\frac{1}{2^8}} = \begin{pmatrix}
        1 & 0 & 0 & 0\\
        0 & 1 & 0 & 0\\
        -1 & 0 & 1 & 0\\
        0 & 0 & 0 & 1
    \end{pmatrix}\\
    M_{\varphi=\infty} &= \left( M_{\varphi=0}M_{\varphi=\frac{1}{2^8}}\right)^{-1} = \begin{pmatrix}
        1 & 1 & -8 & 8\\
        0 & 1 & -8 & 16\\
        1 & 1 & -7 & 8\\
        0 & 0 & -1 & 1
    \end{pmatrix}
    \end{split}
\end{align}
% \begin{equation}
%     M_0 = \begin{pmatrix}
%         1 & -1 & 8 & 8\\
%         0 & 1 & -8 & -16\\
%         0 & 0 & 1 & 0\\
%         0 & 0 & 1 & 1
%     \end{pmatrix},
%     M_{\frac{1}{2^8}} = \begin{pmatrix}
%         1 & 0 & 0 & 0\\
%         0 & 1 & 0 & 0\\
%         -1 & 0 & 1 & 0\\
%         0 & 0 & 0 & 1
%     \end{pmatrix},
%     M_{\infty} = M_{\frac{1}{2^8}}^{-1} M_0^{-1} = \begin{pmatrix}
%         1 & 1 & -8 & 8\\
%         0 & 1 & -8 & 16\\
%         1 & 1 & -7 & 8\\
%         0 & 0 & -1 & 1
%     \end{pmatrix}.
% \end{equation}
Although $\varphi$ is the most natural variable for AESZ 3 and the mirror of the intersection of four quadrics in $\mathbb{CP}^7$, in order to make contact with the fibred product of elliptic surfaces in \eqref{eq: equation for Wtilde}, it is more natural to use the the variable $\phi$, where $\varphi=\left(\frac{\phi}{2^8}\right)^2$. The pullback of AESZ 3 defines a new operator
\begin{equation}
    \mathcal{L}_{\text{AESZ3}}^*=(\phi^2-2^8)\theta_\phi^4+2^2\phi^2\theta_\phi^3+2\cdot3\phi^2\theta_\phi^2+2^2\phi^2\theta_\phi+\phi^2.
\end{equation}
with Riemann symbol
\begin{equation}
    \mathcal{P}\left\{\begin{array}{c c c} 0 & \pm 4^2 & \infty\\
    \hline
    0 & 0 & 1\\
    0 & 1 & 1\\
    0 & 1 & 1\\
    0 & 2 & 1
    \end{array}~; \phi\right\}~.
\end{equation}
 We pullback the integral symplectic basis of solutions in \eqref{eq: rho matrix for AESZ 3} and define a basis of solutions of $\mathcal{L}_{\text{AESZ3}}^*\Pi^*(\phi)=0$ with integral symplectic monodromy by
\begin{equation}
    \Pi^*(\phi)=\rho\begin{pmatrix} 
    1 & 0 & 0 & 0\\
    2\log\frac{1}{2^8} & 2 & 0 & 0\\
    2^2\log^2\frac{1}{2^8} & 2^2 \log\frac{1}{2^8} & 2^2 & 0\\
    2^3 \log^3\frac{1}{2^8} & 2^3\cdot 3 \log^2\frac{1}{2^8} &  2^3\cdot 3\log\frac{1}{2^8} & 2^3
    \end{pmatrix}\varpi^*(\phi),
\end{equation}
where a Frobenius basis of solutions of $\mathcal{L}_{\text{AESZ3}}^*$ is defined as in \eqref{eq: Frobenius basis at MUM point of fourth order PF equation} and $\rho$ is defined in \eqref{eq: rho matrix for AESZ 3}. The power series defining the Frobenius basis of solutions $\varpi^*$ are now given by
\begin{align}
    \label{eq: F_j for pullback of AESZ3}
    \begin{split}
            F_0^*(\phi)&=\sum_{n=0}^\infty \binom{2n}{n}^4\left(\frac{\phi}{2^8}\right)^{2n}\\
            &=1+\frac{\phi ^2}{4096}+\frac{81 \phi ^4}{268435456}+\frac{625 \phi ^6}{1099511627776}+\frac{1500625 \phi ^8}{1152921504606846976}+\ldots\\
             F_1^*(\phi)&=\frac{\phi ^2}{2048}+\frac{189 \phi
   ^4}{268435456}+\frac{4625 \phi ^6}{3298534883328}+\frac{22852375 \phi
   ^8}{6917529027641081856}+\ldots\\
   F_2^*(\phi)&=\frac {\phi ^2} {4096}+\frac {729 \phi
       ^4} {1073741824}+\frac {63925 \phi ^6} {39582418599936}+\frac {687049825 \phi
       ^8} {166020696663385964544}+\ldots\\
       F_3^*(\phi)&=-\frac {3\phi^2} {4096}-\frac {1611\phi^4} {2147483648}-\frac {80735\phi^6} {79164837199872}+\ldots~.
    \end{split}
\end{align}
By analytically continuing $\Pi^*$ from a base point in the upper half $\phi$-plane around the various singularities in an anti-clockwise fashion, we find the following monodromy matrices.
\begin{align}
\label{eq: monodromy matrices for pullback of AESZ3}
    \begin{split}
        M_{\phi=-4^2}&=\left(
\begin{array}{cccc}
 -7 & -8 & 64 & -64 \\
 8 & 9 & -64 & 64 \\
 -1 & -1 & 9 & -8 \\
 -1 & -1 & 8 & -7 \\
\end{array}
\right),
M_{\phi=0}=\left(
\begin{array}{cccc}
 1 & -2 & 32 & 32 \\
 0 & 1 & -32 & -32 \\
 0 & 0 & 1 & 0 \\
 0 & 0 & 2 & 1 \\
\end{array}
\right)\\
M_{\phi=4^2}&=\left(
\begin{array}{cccc}
 1 & 0 & 0 & 0 \\
 0 & 1 & 0 & 0 \\
 -1 & 0 & 1 & 0 \\
 0 & 0 & 0 & 1 \\
\end{array}
\right)\\
M_{\phi=\infty}&=(M_{\phi=-4^2}M_{\phi=0}M_{\phi=4^2})^{-1}=\left(
\begin{array}{cccc}
 -7 & -6 & 32 & -32 \\
 -8 & -7 & 32 & -32 \\
 -6 & -5 & 25 & -24 \\
 -1 & -1 & 6 & -7 \\
\end{array}
\right).
\end{split}
\end{align}
Both $\mathcal{L}_\text{AESZ3}$ and its pull-back $\mathcal{L}_\text{AESZ3}^*$ are invariant under the action of an involution. The integral symplectic solutions $\Pi$ of AESZ3 satisfy
\begin{equation}
    \frac{1}{\sqrt{2^{16}\varphi}}\Pi\left(\frac{1}{2^{16}\varphi}\right)=\left(
\begin{array}{cccc}
 \frac{1}{16} & \frac{1}{16} & -\frac{1}{2} & \frac{1}{2} \\
 0 & -\frac{1}{16} & \frac{1}{2} & -1 \\
 \frac{1}{32} & \frac{1}{64} & -\frac{1}{16} & 0 \\
 \frac{1}{64} & \frac{1}{64} & -\frac{1}{16} & \frac{1}{16} \\
\end{array}
\right)
\Pi(\varphi).
\end{equation}
This identity was computed for real $\varphi$ close to the fixed point $-\frac{1}{2^8}$ and a path connecting $\varphi$ and $\frac{1}{2^{16}\varphi}$ along the upper half $\varphi$-plane. A homotopy inequivalent path would multiply the above matrix by a monodromy matrix. Similarly, we find that
\begin{equation}
  \frac{2^8}{\phi}  \Pi^*\left(\frac{2^8}{\phi}\right) = \left(
\begin{array}{cccc}
 -16 & 0 & 0 & 0 \\
 0 & 16 & 0 & 0 \\
 8 & 4 & -16 & 0 \\
 -4 & 0 & 0 & 16 \\
\end{array}
\right)
\Pi^*(\phi),
\end{equation}
where the above identity was computed near the fixed point $\phi=4^2$ and along a path on the upper half $\phi$-plane connecting $\phi$ and $\frac{2^8}{\phi}$. 

Note that it is not possible make the matrices on the right hand side of the above identities integral symplectic by an overall scaling of the matrices. This is a reflection of the fact that the MUM points of AESZ 3 at $\varphi=0$ and $\varphi=\infty$ are inequivalent in the sense that they are associated with different mirror geometries.\footnote{We are grateful to Thorsten Schimannek for clarifying this point.}\cite{Katz:2023zan} However, at the level of periods, the Frobenius solutions around $\varphi=0$ and $\varphi=\infty$ are identical up to multiplication by an algebraic function of $\varphi$, which leads to the above identities.

\subsection{Picard-Fuchs equation of an associated family of elliptic curves}
In this section, we compute the periods of the holomorphic 1-form for the family of elliptic curves defined in \eqref{eq: equation for the family of elliptic curves E_s}. The family of curves $\mathcal{E}_s$ may be brought into the standard Legendre form
\begin{equation}
    y^2=x(x-1)\left(x-\left(1-\frac{s^2}{4^2}\right)\right)
\end{equation}
by the change of variables
\begin{equation}
    (Y_1,Y_2)=\left(\frac{xs+4y}{xs-4y},\frac{x(1-x)}{y}\right).
\end{equation}
We immediately see that the elliptic curve is singular at $s=0$, $s=\pm 4$ and $s=\infty$.

The holomorphic 1-form on an elliptic fibre is given by
\begin{equation}
    \frac{dx}{2y(s)}=-\frac{dY_1}{\sqrt{Y_1^4+\left(\frac{s^2}{4}-2\right)Y_1^2+1}}.
\end{equation}
Now suppose that $s\in\mathbb{R}$ and $0<s<4$. We define $A\in H_1(\mathcal{E}_s,\mathbb{Z})$ as the homology class of a contour that encircles the branch points 
\begin{equation}
    Y_1=-\frac{1}{2}\sqrt{4-\frac{1}{2}s\left(s-\sqrt{s^2+16}\right)}~~~~\text{and}~~~~Y_1=-\frac{1}{2}\sqrt{4-\frac{1}{2}s\left(s-\sqrt{s^2-16}\right)}
\end{equation}
in an anti-clockwise fashion. Similarly, $B\in H_1(\mathcal{E}_s,\mathbb{Z})$ is defined as the homology class of a contour that encircles the branch points 
\begin{equation}
Y_1=\frac{1}{2}\sqrt{4-\frac{1}{2}s\left(s-\sqrt{s^2-16}\right)}~~~~\text{and}~~~~~Y_1=\frac{1}{2}\sqrt{4-\frac{1}{2}s\left(s+\sqrt{s^2-16}\right)}
\end{equation}
in a clock-wise fashion. It follows that $A\cap B=-B\cap A=1$ and
\begin{equation}
    \label{eq: basis of solutions for pullback of L_A}
     \omega^*(s)=\begin{pmatrix} \int_A \frac{dx}{2y(s)}\\ \int_B \frac{dx}{2y(s)}\end{pmatrix} 
     =
     (2\pi i)\begin{pmatrix}-\frac{1}{2} & ~0 \\ -\frac{\log 2^8}{2 \pi i} & ~2\end{pmatrix}
     \begin{pmatrix} 1 & 0\\ 0 & (2\pi i)^{-1} \end{pmatrix}
     \left(\begin{array}{l}f_0^*(s) \\ f_0^*(s)\log(s) + f_1^*(s)\end{array}\right),
\end{equation}
where
\begin{align}
    \label{eq: f_j for pullback of L_A}
    \begin{split}
   f_0^*(s)&=\sum_{n=0}^\infty \binom{2n}{n}^2\left(\frac{s}{2^4}\right)^{2n}\\
   &=1+\frac{s^2}{64}+\frac{9
   s^4}{16384}+\frac{25
   s^6}{1048576}+\frac{1225
   s^8}{1073741824}+\frac{3969 s^{10}}{68719476736}+\ldots\\
    f^*_1(s)&=\frac{s^2}{64}++\frac{21
   s^4}{32768}+\frac{185
   s^6}{6291456}+\frac{18655
   s^8}{12884901888}+\frac{102501
   s^{10}}{1374389534720}+\ldots~.
    \end{split}
\end{align}
The $A$-integral is holomorphic as $s\rightarrow 0$ and may be computed by first expanding the integrand in $s$, and then evaluating the resulting integrals by computing residues around $Y_1=-1$. Similarly, $B$ is holomorphic as $s\rightarrow 4$ and may be computed around this point as a power series. This solution may then be analytically continued to $s=0$ by noting that the $A$ and $B$ periods are solutions of the Picard-Fuchs equation
\begin{equation}
    (s^2-4^2)\theta_s^2+2s^2\theta_s+s^2
\end{equation}
with Riemann symbol
\begin{equation}
    \label{eq: pullback of L_A}
    \mathcal{P}\left\{\begin{array}{c c c} 0 & \pm 4 & \infty\\
    \hline
    0 & 0 & 1\\
    0 & 0 & 1\\
    \end{array}~; s\right\}~.
\end{equation}
By comparing $f_0^*$ in \eqref{eq: f_j for pullback of L_A} with $F_0^*$ in \eqref{eq: F_j for pullback of AESZ3}, we see that $\mathcal{L}_{\text{AESZ3}}^*$ is the Hadamard product of the above second order differential operator with itself. Note that \eqref{eq: pullback of L_A} is the pullback of the hypergeometric differential operator
\begin{equation}
    (16s'-1)\theta_{s'}^2+16s'\theta_{s'}+4s'
\end{equation}
where $s'=\left(\frac{s}{4^2}\right)^2$. It follows that AESZ 3 is a Hadamard product of this operator with itself.

As expected from geometric considerations (see \eqref{eq: symmetry of E_s}), the solutions $\omega(s)$ have the symmetry
\begin{equation}
\label{eq: s->2^2/s symmetr on periods}
    \frac{4^2}{s}\omega^*\left(\frac{4^2}{s}\right)=4\begin{pmatrix} 1 & 1\\0 &1\end{pmatrix}\omega^*(s)~.
\end{equation}
Note that in computing the above matrix, we need to arbitrarily choose a path between $s$ and $\frac{4^2}{s}$. Thus, the above matrix is only defined up to multiplication by a monodromy matrix. The matrix in \eqref{eq: s->2^2/s symmetr on periods} was computed in a small neighbourhood of the fixed point $s=4$, using a path along the upper half $s$-plane. Also note that this matrix is not a monodromy matrix of $\omega(s)$, because it has a $1$ as the upper off-diagonal entry. The monodromy matrices, for a choice of base point in the upper half $s$-plane and anti-clockwise loops around the various singularities, are generated by
\begin{equation}
    m_{s=-4}=\begin{pmatrix} 5 & 2 \\ -8 & -3\end{pmatrix},~~~m_{s=0} = \begin{pmatrix} 1 & 0 \\ -4 & 1 \end{pmatrix},~~~\text{and}~~~m_{s=4} = \begin{pmatrix} 1 & 2\\ 0 & 1\end{pmatrix}~.
\end{equation}

\section{AESZ 101}
\label{section: AESZ 101}

\subsection{Family of Calabi-Yau threefolds}
Consider the family of elliptic curves in $\mathbb{CP}^2$, defined by the cubic
\begin{equation}
\label{eq: elliptic curve for L_b}
\mathcal{E}_s: x(x-z)(y-z)-s yz(x-y)=0
\end{equation}
with parameter $s$. $\mathcal{E}_s$ is smooth away from $s\in\{0,\frac{1}{2}\left(-11+\pm5\sqrt{5}\right),\infty\}$.

It was shown in \cite{DissertationSamol}, using the results of \cite{2008arXiv0802.3760K}, that the fibred product of the families $\mathcal{E}_s$ and $\mathcal{E}_{\frac{\varphi}{s}}$ may be resolved to a smooth Calabi-Yau threefold with Hodge numbers $h^{1,1}=51$ and $h^{2,1}=1$ (and Euler chracteristic $2\times(51-1)=100$) for generic values of $\varphi$. More precisely, we consider the singular variety in $\mathbb{CP}^2\times\mathbb{CP}^2\times\mathbb{CP}^1$, defined by
\begin{align}
\label{eq: fibred product for AESZ101}
    \begin{split}
        x_1(x_1-z_1)(y_1-z_1)-s y_1z_1(x_1-y_1)&=0\\
        x_2(x_2-z_2)(y_2-z_2)-\left(\frac{\varphi}{s}\right) y_2z_2(x_2-y_2)&=0
    \end{split}
\end{align}
where $[x_i,y_i,z_i]$ are homogeneous coordinates on $\mathbb{CP}^2$ and $s$ is an inhomogeneous coordinate on $\mathbb{CP}^1$. $\varphi$ is a free-parameter that determines complex structure of a smooth Calabi-Yau threefold away from $\varphi\in\{-1,0,\left(\frac{1}{2}\left(-11\pm 5\sqrt{5}\right)\right)^2,\infty\}$. If we consider points on the complex $s$-plane where an elliptic fibre is singular, these values of $\varphi$ are the values for which two singular points collide. A projective small resolution of singularities exists away these values of $\varphi$.

Note that the elliptic curve $\mathcal{E}_s$ is equivalent to the curve $\mathcal{E}_{-\frac{1}{s}}$. This can be seen by making the linear change of variables
\begin{equation}
    \label{eq: involution of L_b on elliptic curve}
    (x,y,z)\mapsto (-z,~x-z,~y-z)~.
\end{equation}
By applying this to both elliptic fibres of \eqref{eq: fibred product for AESZ101}, we see that the Calabi-Yau variety at $\varphi$ is equivalent to that at $\frac{1}{\varphi}$.

We will need the $A$-model topological data (triple intersection number, second Chern class and Euler characteristic) in order to fix an integral symplectic basis of periods around the MUM point $\varphi=0$. One approach is to construct the mirror of the resolution of the variety defined in  \eqref{eq: fibred product for AESZ101}. A natural candidate for the mirror appears in \cite{2010arXiv1006.0223K,2017arXiv170700534B,2017arXiv170609952O,Knapp:2019cih} that is constructed as the intersection of two Grassmannians in $\mathbb{CP}^9$. More precisely, the Pl{\"u}cker embedding defines a map from the Grassmannian $Gr(2,5)$ (of two dimensional linear subspaces of $\mathbb{C}^5$) to the projective space $\mathbb{CP}^9$. $GL(10,\mathbb{C})$ acts in the obvious way on the homogeneous coordinates of $\mathbb{CP}^9$ and, by composing the Plucker embedding with a $GL(10,\mathbb{C})$ transformation, one may generate many embeddings of $Gr(2,5)$ in $\mathbb{CP}^9$. For generic choices of $GL(10,\mathbb{C})$ transformation, the intersection of the original and a transformed copy of $Gr(2,5)$ in $\mathbb{CP}^9$ will be a smooth Calabi-Yau threefold with Hodge numbers $h^{1,1}=1$ and $h^{2,1}=51$. This Calabi-Yau threefold has the triple intersection number $H^3=25$, second Chern class $c_2.H=70$ and Euler characteristic $\chi=-100$.\cite{2010arXiv1006.0223K}

An alternative approach for computing the A-model topological data is to first compute the Picard-Fuchs equation associated with \eqref{eq: fibred product for AESZ101}, and then use monodromy to constrain the topological data of the mirror, as was done in \cite{2004math.....12539V}. This approach proceeds as follows. We know that the Picard-Fuchs equation must be the Hadamard product of two second order Picard-Fuchs equations associated with families of elliptic curves $\mathcal{E}_s$ and $\mathcal{E}_\frac{\varphi}{s}$, so it is straightforward to find on a computer (by writing down the holomorphic solution and searching for a fourth order differential operator that annihilates it). In this way, we find the Picard-Fuchs equation AESZ 101 (see following subsection for the equation and its solutions). We compute the solutions of the Picard-Fuchs equation in a suitable basis and assume that a matrix like the one in \eqref{eq: rho matrix for AESZ 3} determines a basis of periods with integral symplectic monodromy. We then numerically compute the monodromy matrices and require that they be symplectic with integer coefficients. Since these matrices must have integer coefficients, a single decimal place of precision is sufficient. We also require that the monodromy around the nearest conifold take a standard form with a single off-diagonal component. In this way, we find that $H^3=25$ and $c_2\cdot H=70$ are the only choices consistent with $\chi=-100$.

%--------------------------------------------------------------------------------------------------------
\subsection{Solutions of AESZ 101}
%--------------------------------------------------------------------------------------------------------
AESZ 101 is the operator
\begin{align}
    \label{eq: AESZ 101}
    \begin{split}
        \mathcal{L}_{\text{AESZ101}}=&(\varphi -1)^2 (\varphi +1) \left(\varphi ^2-123 \varphi +1\right)\theta_\varphi^4\\
        &+2 (\varphi -1) \varphi  \left(2 \varphi ^3-125 \varphi ^2+244 \varphi
   +121\right)\theta_\varphi^3\\
   &+\varphi  \left(6 \varphi ^4-205 \varphi ^3+6/89 \varphi ^2-787 \varphi
   -187\right)\theta_\varphi^2\\
   &+2 \varphi  \left(2 \varphi ^4-39 \varphi ^3+105 \varphi ^2-277 \varphi
   -33\right)\theta_\varphi\\
   &+\varphi  \left(\varphi ^4-12 \varphi ^3+12 \varphi ^2-124 \varphi
   -9\right)
    \end{split}
\end{align}
where $\theta_{\varphi}=\varphi\frac{d}{d\varphi}$. AESZ 101 has the Riemann symbol
\begin{equation}
    \label{eq: Riemann symbol of AESZ101}
    \mathcal{P}\left\{\begin{array}{c c c c c} -1 & 0 & s_\pm^2 & 1 & \infty\\
    \hline
    0 & 0 & 0 & 0 & 1\\
    1 & 0 & 1 & 1 & 1\\
    1 & 0 & 1 & 3 & 1\\
    2 & 0 & 2 & 4 & 1
    \end{array}~; \varphi\right\}~,
\end{equation}
where $s_+$ and $s_-$ are the positive and negative solutions of $s_\pm^2+11s_\pm-1=0$ respectively i.e. $s_\pm=\frac{1}{2}\left(-11\pm5\sqrt{5}\right)$. A Frobenius basis of solutions of around the MUM at $\varphi=0$ is defined as in \eqref{eq: Frobenius basis at MUM point of fourth order PF equation} and the relevant power series are now given by
\begin{align}
\label{eq: F_j for AESZ101}
    \begin{split}
        F_0(\varphi)&=\sum_{n=0}^{\infty}\left(\sum_{k=0}^n \binom{n}{k}^2\binom{n+k}{k}\right)^2\varphi^n\\
        &=1+9 \varphi +361 \varphi ^2+21609 \varphi^3+1565001 \varphi ^4+126630009 \varphi ^5+\ldots\\
        F_1(\varphi)&=30 \varphi+1425 \varphi ^2+90895 \varphi ^3+\frac{13604625
        \varphi ^4}{2}+\frac{1123000637 \varphi ^5}{2}+\ldots\\
        F_2(\varphi)&=26 \varphi+\frac{4865 \varphi^2}{2}+\frac{3369877 \varphi ^3}{18}+\frac{122552385\varphi ^4}{8}+\frac{2398214876497 \varphi ^5}{1800}+\ldots\\
        F_3(\varphi)&=-156 \varphi-\frac{10599 \varphi^2}{2}-\frac{4192853 \varphi ^3}{18}-\frac{195151095\varphi ^4}{16}+\ldots~.
    \end{split}
\end{align}
A basis of periods $\Pi$ with integral symplectic mononodromy is defined as in \eqref{eq: Pi = rho*varpi}. We use the matrix $\rho$ in \eqref{eq: rho matrix for AESZ 3} with the topological data $H^3=25$, $c_2\cdot H=70$ and $\chi=-100$. We choose a base point in the upper half $\varphi$-plane and anti-clockwise loops around the various singularities. Alternatively, this base point may be deformed to the real $\varphi$ axis (this will be helpful for the discussion in Appendix~\ref{sec: varphi-monodromy as a braid}). Both choices of base point lead to the same monodromy matrices, which are given by
\begin{align}
    \label{eq: monodromy matrices for AESZ 101}
    \begin{split}
    M_{\varphi=-1}&=\left(
        \begin{array}{cccc}
            -19 & -10 & 50 & -20 \\
            8 & 5 & -20 & 8 \\
            -8 & -4 & 21 & -8 \\
            -4 & -2 & 10 & -3 \\
        \end{array}
        \right),~
    M_{\varphi=0}=\left(
        \begin{array}{cccc}
            1 & -1 & 10 & 13 \\
            0 & 1 & -12 & -25 \\
            0 & 0 & 1 & 0 \\
            0 & 0 & 1 & 1 \\
        \end{array}
        \right)\\
    M_{\varphi=s_+^2}&=\left(
        \begin{array}{cccc}
            1 & 0 & 0 & 0 \\
            0 & 1 & 0 & 0 \\
            -1 & 0 & 1 & 0 \\
            0 & 0 & 0 & 1 \\
        \end{array}
        \right),
    M_{\varphi=s_-^2}=\left(
        \begin{array}{cccc}
            1 & 0 & 0 & 0 \\
            -60 & 1 & 0 & 225 \\
            -16 & 0 & 1 & 60 \\
            0 & 0 & 0 & 1 \\
        \end{array}
        \right)\\
    M_{\varphi=\infty} &=(M_{\varphi=-1}M_{\varphi=0}M_{\varphi=s_+^2}M_{\varphi=s_-^2})^{-1}=
    \left(
        \begin{array}{cccc}
            -19 & -9 & 40 & -8 \\
            -252 & -95 & 392 & 208 \\
            -75 & -29 & 121 & 52 \\
            -4 & -2 & 9 & -3 \\
        \end{array}
        \right)
    \end{split}
\end{align}
As an aside, note that, although $\varphi=1$ is a singularity of the Picard-Fuchs equation, the underlying Calabi-Yau variety is smooth and the periods have trivial monodromy around this point.\footnote{The determinant of the Wronskian matrix vanishes at $\varphi=1$. It is an apparent singularity.} This follows from the lack of repeated indices in the Riemann symbol \eqref{eq: Riemann symbol of AESZ101} at $\varphi=1$. This point is quite interesting in its own right. As argued in the previous subsection, the Calabi-Yau variety at $\varphi$ is equivalent to that at $\frac{1}{\varphi}$. This should be visible in the periods $\Pi(\varphi)$ and, indeed, we find that
\begin{equation}
    \label{eq: involution of AESZ 3}
    \frac{1}{\varphi}\Pi\left(\frac{1}{\varphi}\right)=\left(
\begin{array}{cccc}
 -4 & 0 & 0 & 15 \\
 7 & 4 & -15 & 0 \\
 0 & 1 & -4 & 7 \\
 -1 & 0 & 0 & 4 \\
\end{array}
\right)\Pi(\varphi)
\end{equation}
where the matrix on the right hand side has been computed in a small neighbourhood of the fixed point $\varphi=1$. The matrix on the right hand side of the above identity is an element of $Sp(4,\mathbb{Z})$ and can be understood as a map from the cohomology of the underlying smooth Calabi-Yau $X_\varphi$ to that of $X_\frac{1}{\varphi}$. It has the eigenvalues $(+1,+1,-1,-1)$ and, at the smooth fixed point $\varphi=1$, it splits $H^3(X_1,\mathbb{Q})$ into positive and negative eigenspaces. Elements in the positive eigenspace will Hodge numbers $(3,0)+(0,3)$ whereas those in the negative eigenspace will have Hodge numbers $(2,1)+(1,2)$. Such points are known as attractor points of rank two, which are named as such because $H^{3,0}(X_1)\oplus H^{0,3}(X_1)$ is the complexification of a rank two sublattice of $H^3(X,\mathbb{Z})$. A numerical computation reveals that
\begin{equation}
    \Pi(1)=\eta_1\begin{pmatrix}6 \\ 1 \\ 3 \\ 2 \end{pmatrix} + \eta_2\begin{pmatrix}0 \\ 5 \\ 1 \\ 0 \end{pmatrix},
\end{equation}
where
\begin{align}
    \begin{split}
        \eta_1=&~~28.985700535271822856288566710613439081713073465729\ldots\\
        \eta_2=&-i89.4161554867742890655454362164259781884639062623\ldots~.
    \end{split}
\end{align}
Recall that the covariant derivative of special geometry acts on the holomorphic $3$-form $\Omega$ (and its periods) as $(\partial_\varphi+\partial_\varphi K)\Omega$. As a cohomology class, $D_\varphi\Omega \in H^{2,1}(X)$ and the K\"{a}hler potential $K$ is defined as
\begin{equation}
    K=-\log\left(-i\int_X\Omega\wedge\overline{\Omega}\right)=-\log\left(-i\Pi^\dagger\Sigma_4 \Pi\right),
\end{equation}
where 
\begin{equation}
    \Sigma_4=\begin{pmatrix} 0 & \mathds{1}_2\\- \mathds{1}_2 & 0\end{pmatrix}
\end{equation}
is the standard $4\times 4$ symplectic form. We find that
\begin{equation}
D_{\varphi}\Pi(1)=\eta_3\begin{pmatrix}10 \\ 1 \\ 5 \\ 2\end{pmatrix}+\eta_4\begin{pmatrix}0 \\ 3 \\ 1 \\ 0 \end{pmatrix}
\end{equation}
where
\begin{align}
    \begin{split}
        \eta_3=&~ 4.5830075667086921421687455172516021305804602797484\ldots\\
        \eta_4=&~ -i9.96834656548114388389121094203891689421621590045\ldots~.
    \end{split}
\end{align}
The transcendental numbers $\eta_i$ are known to be critical values of $L$-functions associated to the Calabi-Yau threefold $X_{\varphi=1}$ (up to rational multiples and factors of $(2\pi i)$). In other words, they contain information about the arithmetic geometry of the Calabi-Yau at $\varphi=1$. We will not have much more to say about such points in this paper. The interested reader is referred to \cite{Candelas:2019llw,Bonisch:2022mgw}.

\subsection{Picard-Fuchs equation of an associated family of elliptic curves}

The elliptic curve in \eqref{eq: elliptic curve for L_b} is defined by the polynomial
\begin{equation}
    P(x,y,z,s)=x(x-z)(y-z)-s y z(x-y)
\end{equation}
where $x$, $y$ and $z$ define homogeneous coordinates on $\mathbb{CP}^1$ and $s$ is a complex parameter. On the $z=1$ coordinate patch, the holomorphic 1-form may be taken to be
\begin{equation}
    \frac{dx}{\partial_y P|_{P=0}}=\frac{dx}{\sqrt{x(-4s+x+s(6+s)x-2(1+s)x^2+x^3)}}~.
\end{equation}
We define a basis of $A$ and $B$ periods as follows. The branch points are the solutions of
\begin{equation}
    x(-4s+x+s(6+s)x-2(1+s)x^2+x^3)=0.
\end{equation}
For $s\in\mathbb{R}$ and $0<s<\frac{1}{2}\left(-11+5\sqrt{5}\right)$, the branch points are all real and are ordered from left to right on the real line. We define $A\in H_1(\mathcal{E}_s,\mathbb{Z})$ as the homology class of a contour encircling the first and second branch points in an anti-clockwise fashion. Similarly, $B\in H_1(\mathcal{E}_s,\mathbb{Z})$ is defined as the homology class of a contour encircling the second and third branch points in an anti-clockwise fashion. This implies that
\begin{equation}\label{eq: integral symplectic periods for L_b}
\omega(s)=
\begin{pmatrix}\int_A\frac{dx}{\partial_y P|_{P=0}} \\ \int_B \frac{dx}{\partial_y P|_{P=0}}\end{pmatrix}
=(2\pi i)
\begin{pmatrix} 1 & 0 \\ 0 & 5 \end{pmatrix}
\begin{pmatrix} 1  & 0 \\ 0 & \left(2\pi i\right)^{-1} \end{pmatrix} 
\left(\begin{array}{l} f_0(s) \\ f_0(s) \log(s) + f_1(s)\end{array}\right)
\end{equation}
where
%\begin{subequations}%\label{eq:y}
\begin{align}
\label{eq: f_j for L_b}
\begin{split}
%\label{eq:y:1}
f_0(s)  & = \sum_{n=0}^\infty\left(\sum_{k=0}^{n}\binom{n}{k}^2\binom{n+k}{k}\right)s^n\\
        & =1 + 3s + 19s^2 + 147s^3 + 1251s^4 + 11253s^5 + 104959s^6 + 1004307s^7 + \ldots\\
f_1(s)  & = 5s + \frac{75}{2}s^2 + \frac{1855}{6}s^3 + \frac{10875}{4}s^4 + \frac{299387}{12}s^5 + \frac{943397}{4}s^6 +   \ldots~.
\end{split}
\end{align}
The $A$ period may be computed by expanding the denominator of the integrand around $s=0$ and evaluating the resulting integrals by computing residues. The $B$ integral is holomorphic around the point $s=\frac{1}{2}\left(-11+5\sqrt{5}\right)$ and may be similarly computed around this point. We then analytically continue it back to zero by noting that the $A$ period is a annihilated by the differential operator
\begin{equation}
    \label{eq: L_b}
    (s^2+11s-1)\theta_s^2+s(2s+11)\theta_s+s(s+3)
\end{equation}
with the Riemann symbol
\begin{equation}
    \mathcal{P}\left\{\begin{array}{c c c} 0 & \pm s_\pm & \infty\\
    \hline
    0 & 0 & 1\\
    0 & 0 & 1\\
    \end{array}~; s\right\},
\end{equation}
where $s_\pm=\frac{1}{2}\left(-11\pm5\sqrt{5}\right)$. The $B$ period must be annihilated by the same operator, so we numerically evaluate the logarithmic solution of the above differential equation around $s=0$ and evaluate it for some real $0<s<s_+$. By numerically evaluating the $B$ period computed around $s_+$ at this same value of $s$, we determine the constant $5$ in \eqref{eq: integral symplectic periods for L_b}.

By comparing the holomorphic solutions in \eqref{eq: F_j for AESZ101} and \eqref{eq: f_j for L_b}, we see that AESZ 101 is a Hadamard product of \eqref{eq: L_b} with itself.

Recall that the elliptic curve $\mathcal{E}_s$ is isomorphic to the elliptic curve $\mathcal{E}_{-\frac{1}{s}}$. As was the case in \eqref{eq: involution of AESZ 3}, we see this as a symmetry of the elliptic periods
\begin{equation}
    -\frac{1}{s}\omega\left(-\frac{1}{s}\right)=\begin{pmatrix} -2 & 1\\ -5 & 2\end{pmatrix}\omega(s).
\end{equation}
The above matrix has been computed in a small neighbourhood of the smooth fixed point $s=i$ and, as always, is only defined up to multiplication by a monodromy matrix.\footnote{The smooth fixed points at $s=\pm i$ have complex multiplication as a result of the symmetry of the family.} For a choice of base point in the upper half plane and anti-clockwise loops around the various singularities, we find that the monodromy matrices are generated by
\begin{equation}
    m_{s=s_-}=\begin{pmatrix} 11 &-4\\25 & -9\end{pmatrix},~~~m_{s=0} = \begin{pmatrix} 1 & 0 \\ 5 & 1 \end{pmatrix}~~~\text{and}~~~m_{s=s_+} = \begin{pmatrix} 1 & -1\\ 0 & 1\end{pmatrix}~.
\end{equation}

%--------------------------------------------------------------------------------------------------------------
\section{$3$-Cycles on Fibred Products of Elliptic Surfaces}
%--------------------------------------------------------------------------------------------------------------
\subsection{Solutions of Hadamard product}
Consider a family of elliptic curves $\mathcal{E}\rightarrow \mathbb{CP}^1$, where the elliptic fibre degenerates at some finite number of points $\Delta \subset\mathbb{CP}^1$. We write $\mathcal{E}_s$ for the fibre at some $s\in\mathbb{CP}^1$. In studying Hadamard products, we have been considering fibred products such that the fibre at $s\in\mathbb{CP}^1$ has the form $\mathcal{E}_s\times \mathcal{E}_\frac{\varphi}{s}$ for some parameter $\varphi$ of the resulting $3$-fold $\mathcal{E}\times_{\mathbb{CP}^1}\widetilde{\mathcal{E}}$.\footnote{The two examples of Hadamard products that we consider in this paper are Hadamard products of two identical differential operators. We therefore consider fibred products $\mathcal{E}\times_{\mathbb{CP}^1}\mathcal{\widetilde{E}}$ where the second family of elliptic curves $\widetilde{\mathcal{E}}\rightarrow\mathbb{CP}^1$ is related to the first by a simple change of variables on the base $\mathbb{CP}^1$. Of course, we are free to consider Hadamard products of two different differential operators, and we expect that the main results of this paper will hold for these examples as well.} If the elliptic curves are presented in Weierstrass form, and we remove points on the base where an elliptic fibre degenerates, a holomorphic $3$-form on $\mathcal{E}\times_{\mathbb{CP}^1\backslash \Delta_1\cup\Delta_2}\mathcal{\widetilde{E}}$ is given by
\begin{equation}
    \label{eq: holomorphic 3-form on fibred product of elliptic curves}
    \frac{dx_1}{2y_1(s)}\wedge\frac{dx_2}{2y_2\left(\frac{\varphi}{s}\right)}\wedge\frac{ds}{s}
\end{equation}
where $y_i$ and $x_i$ are the usual variables in the Weierstrass equation of the $i^{th}$ elliptic fibre, $s$ is a coordinate on the the base $\mathbb{CP}^1$ and $\Delta_i$ is the finite set of points on the base where the $i^{th}$ elliptic fibre degenerates. We would like to integrate the above $3$-form over a 3-chain in $\mathcal{E}\times_{\mathbb{CP}^1\backslash \Delta_1\cup\Delta_2}\mathcal{\widetilde{E}}$ so that the integral does not vanish for generic $\varphi$. A $3$-chain with this property must necessarily take the form of a product of two circles, one  in each of the factors of $\mathcal{E}_s\times \mathcal{E}_\frac{\varphi}{s}$, fibred over a contour on $\mathbb{CP}^1\backslash \Delta_1\cup\Delta_2$. We may further insist that the $3$-chain is a $3$-cycle by requiring that the $2$-boundary vanishes. We may arrange this by choosing a non-trivial contour in $\mathbb{CP}^1\backslash \Delta_1\cup\Delta_2$ and a choice of $\gamma_1\otimes\gamma_2$ in $H_1\left(\mathcal{E}_s,\mathbb{Z}\right)\otimes H_1\left(\mathcal{E}_{\frac{\varphi}{s}},\mathbb{Z}\right)$ that is invariant when transported around the contour. This defines a topological $3$-torus. Alternatively, the contour on the base may be open, with end points in $\Delta_1\cup\Delta_2$. The requirement of trivial $2$-boundary is again a constraint on the choice of homology classes in the fibres. Finally, we may consider collections of such contours on the base, each with a choice of homology class in the fibre, such that the boundary contributions cancel out and the collection of $3$-chains becomes a $3$-cycle. We will see examples of all three types of $3$-cycles, but first we fix some notation.

We write the periods of the holomorphic $(3,0)$ form on a Calabi-Yau threefold as
\begin{equation}
    \Pi(\varphi)=\left(\int_{A_I}\Omega(\varphi),\int_{B^I}\Omega(\varphi)\right)^T=\left(F_I(\varphi),X^I(\varphi)\right)^T
\end{equation}
where $A_I$ and $B^I$ generate the torsion-free part of $H_3\left(X,\mathbb{Z}\right)$ and intersect as $A_I\cap B^J=-B^J\cap A_I=\delta_I^J$, with all other intersections vanishing. The index $I$ runs from $I=1$ to $I=b_3(X)=4$. We define a dual basis in cohomology $\{\alpha^I,\beta_J\}$ via the relations
\begin{equation}
    \int_{A_J}\alpha^I=-\int_{B^I}\beta_J=\int_X\alpha^I\wedge\beta_J=\delta_J^I
\end{equation}
with all other integrals vanishing. The holomorphic $(3,0)$-form may now be expressed as
\begin{equation}
    \Omega(\varphi)=F_I(\varphi)\alpha^I-X^I(\varphi)\beta_I~.
\end{equation}
A torsion-free, integral cohomology class
\begin{equation}
p_I\alpha^I-q^I\beta_I
\end{equation}
is Poincar{é} dual to the homology class
\begin{equation}
    -q^IA_I+p_IB^I
\end{equation}
where $p_I,q^J\in \mathbb{Z}$. We will consistently abuse notation and represent both by the integral periods 
\begin{equation}
    \Gamma = \begin{pmatrix} p_I \\ q^I\end{pmatrix}= \begin{pmatrix} \int_{A_J} \left(p_I\alpha^I-q^I\beta_I\right)\\ \int_{B^I} \left(p_I\alpha^I-q^I\beta_I\right)\end{pmatrix}\in\mathbb{Z}^4.
\end{equation}
We will sometimes write $\Pi_\Gamma(\varphi)$ for the integral of the holomorphic $(3,0)$-form over the homology class represented by $\Gamma$ i.e.
\begin{equation}
    \Pi_\Gamma(\varphi)=\Gamma^T\Sigma_4\Pi(\varphi)=\int_{-q^IA_I+p_IB^I}\Omega=\int_X\Omega\wedge\left(p_I\alpha^I-q^I\beta_I\right)
\end{equation}
where
\begin{equation}
    \Sigma_n=\begin{pmatrix} 0 & \mathds{1}_{\frac{n}{2}}\\ -\mathds{1}_{\frac{n}{2}} & 0\end{pmatrix}
\end{equation}
is the standard $n\times n$ symplectic form and $n\in2\mathbb{Z}$.

In precisely the same way, we may compute the integral of the holomorphic $1$-form on an elliptic curve over a $1$-homology class that is Poincar{é} dual to a cohomology class with periods $\gamma\in\mathbb{Z}^2$. We denote this analogously by
\begin{equation}
    \omega_\gamma(s)=\gamma^T\Sigma_2\omega(s)~.
\end{equation}

Suppose that  $N\in\mathbb{Z}$ and consider the network
\begin{equation}
    \label{eq: collection of contours and cycles in the fibres}
    \left\{\left(\ell^{(i)},\gamma_1^{(i)},\gamma_2^{(i)}\right)\right\}_{i=1}^N,
\end{equation}
defined by a collection of contours $\ell^{(i)}$ between two singularities on the $s$-plane and $1$-homology classes $\gamma_1^{(i)}$ and $\gamma_2^{(i)}$ in the first and second elliptic fibre, respectively. The tensor product of the two homology classes is fibred over the corresponding contour $\ell^{(i)}$, and the sum of the resulting $3$-chains is a $3$-cycle i.e. the $2$-boundary must vanish. We may integrate \eqref{eq: holomorphic 3-form on fibred product of elliptic curves} over this $3$-cycle and find that
\begin{equation}
\label{eq: integral over a network equals a period}
\Pi_\Gamma(\varphi)=\sum_{i=1}^{N}\int_{\ell^{(i)}}\omega_{\gamma^{(i)}_1}(s)\omega_{\gamma^{(i)}_2}\left(\frac{\varphi}{s}\right)\frac{ds}{s}
\end{equation}
for some (co)-homology class, represented by $\Gamma\in\mathbb{Z}^4$. This equation shall be the workhorse of the remainder of this paper. It allows us to compute the (co)homology class associated to a network like the one in \eqref{eq: collection of contours and cycles in the fibres}.

When studying integrals like the one in \eqref{eq: integral over a network equals a period}, one must worry about monodromy as $\varphi$ is varied around singular points. Braids provide a useful tool for keeping track of how the contours on the base must vary as $\varphi$ is varied. We explain this in Appendix~\ref{sec: varphi-monodromy as a braid}.

Note that, in the examples described in the previous section, there exists a smooth Calabi-Yau variety $X_\varphi$ (for generic $\varphi$) and a map $X_\varphi\rightarrow \mathcal{E}\times_{\mathbb{CP}^1}\mathcal{\widetilde{E}}$, so $\Gamma$ should be understood as representing some class in $H_3(X_\varphi,\mathbb{Q})$, which is pushed forward onto the sum of $3$-chains in \eqref{eq: collection of contours and cycles in the fibres}. This is because the basis of periods $\Pi(\varphi)$ is determined, via mirror symmetry, for $X_\varphi$.

\subsection{Double cover of AESZ 3}
In section \ref{section: AESZ 3}, we studied a double cover of AESZ 3. Recall that both $\mathcal{L}_\text{AESZ3}$ and the pull-back $\mathcal{L}^*_\text{AESZ3}$ are Hadamard products. Let $\omega^*(s)$ be the solution of $\mathcal{L}^*_\text{AESZ3}\omega^*(s)=0$ with monodromy in $SL(2,\mathbb{Z})$, as defined in Section~\ref{section: AESZ 3}, and consider the monodromy of $\omega^*(s)\otimes\omega^*\left(\frac{\varphi}{s}\right)$ around some singularity $s_*$. The monodromy matrices are of the form $m_{s_*}\otimes\widetilde{m}_{s_*}$. We choose some real $0<\phi<4^2$ and a set of generators of $\pi_1\left(\mathbb{CP}^1\backslash\{-4,-\frac{\phi}{4},0,\frac{\phi}{4},4,\infty\},s_0\right)$ in Figure~\ref{fig: generators of fundamental group of s-plane for AESZ101}. We must also fix the branch that $\omega^*(s)$ and $\omega^*\left(\frac{\varphi}{s}\right)$ start on. We do this by choosing the principal branch of the logarithms in \eqref{eq: basis of solutions for pullback of L_A} and requiring that $\omega^*(s)$ and $\omega^*\left(\frac{\varphi}{s}\right)$ are on the same branch when evaluated on the interval $\frac{\phi}{4}<s<4$. With these conventions, we find the monodromy matrices in Table~\ref{table: monodromy of L_A for omegaxomega}.
\begin{figure}
    % [trim={left bottom right top},clip]
    \begin{tikzpicture}
    \node[anchor=south west,inner sep=0] (image) at (0,0) {\frame{\includegraphics[width=\textwidth,trim={0 19cm 0 1cm},clip]{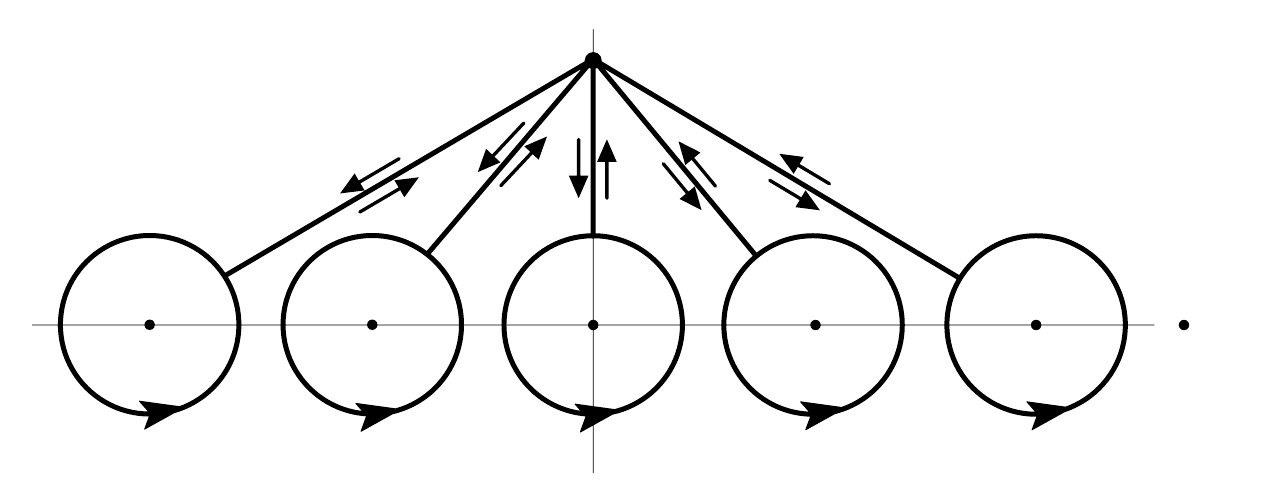}}};
    \begin{scope}[
        x={(image.south east)},
        y={(image.north west)}
    ]
        %\node [white, font=\bfseries] at (0.25,0.65) {$P(X|Y)$};
        \node at (0.126,0.27) {$s=-4$};
        \node at (0.29,0.41) {$\color{red} s=-\frac{\phi}{4}$};
        \node[fill=white] at (0.465,0.27) {$s=0$};
        \node[fill=white] at (0.465,0.41) {$\color{red} s=0$};
        \node at (0.64,0.41) {$\color{red} s=\frac{\phi}{4}$};
        \node at (0.81,0.27) {$s=4$};
        \node at (0.935,0.27) {$s=\infty$};
        \node at (0.935,0.41) {$\color{red} s=\infty$};
        \node at (0.25,0.70) {$\ell_{-4}$};
        \node at (0.25,0.10) {$\ell_{-\frac{\phi}{4}}$};
        \node at (0.42,0.10) {$\ell_{0}$};
        \node at (0.60,0.10) {$\ell_{\frac{\phi}{4}}$};
        \node at (0.66,0.70) {$\ell_{4}$};
        \node at (0.436,0.91) {$s_0$};
    \end{scope}
    \end{tikzpicture}
    \captionsetup{format=hang}
    \caption{Generators of $\pi_1\left(\mathbb{CP}^1\backslash\{-4,-\frac{\phi}{4},0,\frac{\phi}{4},4,\infty\},s_0\right)$ in the complex $s$-plane for real $\phi$ such that $0<\phi<4^2$. We have listed points on the $s$-plane where the first elliptic curve degenerates below the real axis (in black) and points where the second elliptic curve degenerates above the real axis (in red). We choose a base point $s_0$ with very large imaginary part.}
    \label{fig: generators of fundamental group of s-plane for pullback of AESZ 3}
\end{figure}
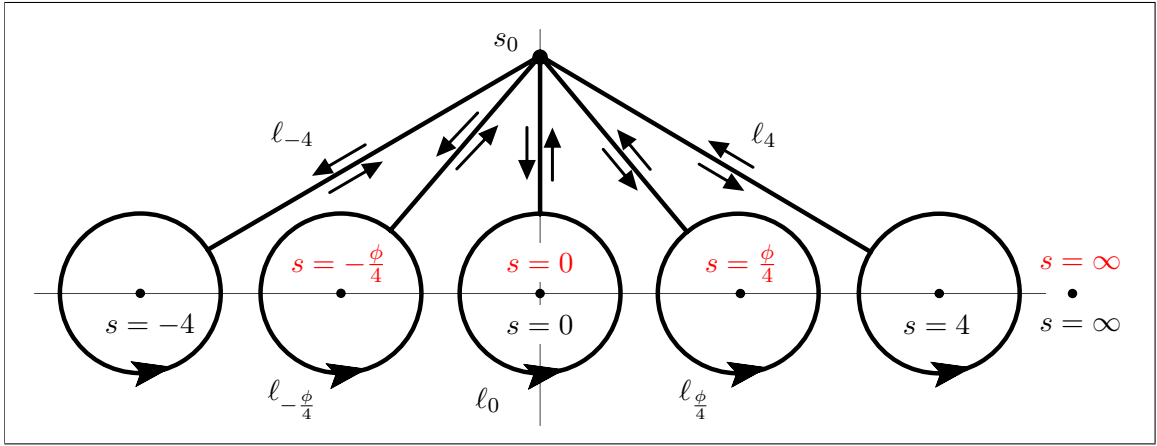

In addition to the monodromy matrices, we also determine the vanishing cycles associated with the various singularities in Table~\ref{table: monodromy of L_A for omegaxomega}. These are listed as integer vectors $v_{s_*}$ and $\widetilde{v}_{s_*}$, which are the periods of the cohomology class Poincar{é} dual to the vanishing cycle. They may be read off from the monodromy matrices in Picard-Lefschetz form
\begin{equation}
    m_{s_*}=\mathds{1}_2+c_{s_*} v_{s_*}\left(\Sigma_2 v_{s_*}\right)^T~~~~~~~\text{and}~~~~~~~\widetilde{m}_{s_*}=\mathds{1}_2+\widetilde{c}_{s_*} \widetilde{v}_{s_*}\left(\Sigma_2 \widetilde{v}_{s_*}\right)^T.
\end{equation}
\begin{table}[h]
\centering
\begin{tabular}{|c||c|c|c|c|c|c|}
\hline
$s_*$ & $m_{s_*}$ & $c_{s_*}$ & $v_{s_*}$ & $\widetilde{m}_{s_*}$ & $\widetilde{c}_{s_*}$ & $\widetilde{v}_{s_*}$ \\
\hline
\hline
&&&&&&\\[-9pt]
%\hhline{=======}
$-4$ & $\left(\begin{matrix} 5 & 2 \\ -8 & -3 \end{matrix}\right)$ & $-2$ & $\begin{pmatrix} -1 \\ 2\end{pmatrix}$ & $\mathds{1}_2$ & n/a & n/a \\[15pt]
\hline
&&&&&&\\[-9pt]
$-\frac{\phi}{4}$ & $\mathds{1}_2$ & n/a & n/a & $\begin{pmatrix} -3 & 2 \\ -8 & 5\end{pmatrix}$ & $-2$ & $\begin{pmatrix} 1 \\ 2\end{pmatrix}$ \\[15pt]
\hline
&&&&&&\\[-9pt]
0 & $\begin{pmatrix} 1 & 0\\ -4 & 1\end{pmatrix}$ & $-4$ & $\begin{pmatrix} 0 \\1\end{pmatrix}$ & $\begin{pmatrix} -3 & 4\\ -4 & 5\end{pmatrix}$ & $-4$ & $\begin{pmatrix}1 \\1\end{pmatrix}$ \\[15pt]
\hline
&&&&&&\\[-9pt]
$\frac{\phi}{4}$ & $\mathds{1}_2$ & n/a & n/a & $\begin{pmatrix} 1 & 2\\ 0 & 1\end{pmatrix}$ & $-2$ & $\begin{pmatrix} 1 \\ 0\end{pmatrix}$ \\[15pt]
\hline
&&&&&&\\[-9pt]
$4$& $\begin{pmatrix} 1 & 2\\ 0 & 1\end{pmatrix}$ & $-2$ & $\begin{pmatrix} 1 \\ 0 \end{pmatrix}$ & $\mathds{1}_2$ & n/a & n/a \\[15pt]
\hline
&&&&&&\\[-9pt]
$\infty$ & $\begin{pmatrix} 5 & 4\\ -4 & -3\end{pmatrix}$ & $-4$ & $\begin{pmatrix} -1 \\ 1\end{pmatrix}$ & $\begin{pmatrix} 1 & 0 \\ -4 & 1\end{pmatrix}$ & $-4$ & $\begin{pmatrix}0 \\ 1\end{pmatrix}$ \\[15pt]
\hline
\end{tabular}
\captionsetup{format=hang}
\caption{The monodromy matrices of the elliptic periods associated with the Hadamard product $\mathcal{L}_{\text{AESZ3}}^*$.}
\label{table: monodromy of L_A for omegaxomega}
\end{table}
With the monodromy of $\omega^*(s)\otimes\omega^*\left(\frac{\phi}{s}\right)$ in hand, we may now begin identifying $3$-cycles. The first $3$-cycle we consider is the $T^3$ that is responsible for the Hadamard product structure of the fourth order Picard-Fuchs equation. We start by re-expressing \eqref{eq: F_j for pullback of AESZ3} as
\begin{equation}
    \frac{1}{4}\left(1,0,0,0\right)\Sigma_4\Pi^*(\phi)=\frac{(2\pi i)^3}{4}\sum_{n=0}^\infty \binom{2n}{n}^4\left(\frac{\phi}{2^8}\right)^{2n}.
\end{equation}
Note that $(0,1)^T\otimes (0,1)^T$ is invariant under monodromy around $\ell_{-\frac{\phi}{4}}\ell_0\ell_{\frac{\phi}{4}}$. The union of an $S_1\times S_1$ in the homology class $\left(0,1\right)^T\otimes\left(0,1\right)^T$ fibred over a loop in the homotopy class $\ell_{-\frac{\phi}{4}}\ell_0\ell_\frac{\phi}{4}$ determines a $3$-cycle. For $0<|\phi|<4^2$, we may compute
\begin{alignat}{2}
    \nonumber
    \oint_{\ell_{-\frac{\phi}{4}}\ell_0\ell_{\frac{\phi}{4}}}\omega^*_{\begin{psmallmatrix} 0 \\ 1\\ \end{psmallmatrix}}(s)\omega^*_{\begin{psmallmatrix} 0 \\ 1\\ \end{psmallmatrix}}\left(\frac{\phi}{s}\right)\frac{ds}{s} =& \frac{(2\pi i)^2}{4}\sum_{m,n=0}^{\infty}&&\Bigg(\binom{2m}{m}^2\binom{2n}{n}^2\left(\frac{1}{2^8}\right)^{2m+2n}\\
    & && \times \phi^{2m}\oint_{\ell_{-\frac{\phi}{4}}\ell_0\ell_{\frac{\phi}{4}}}s^{2m-2n}
    \frac{ds}{s}\Bigg)\\
    \nonumber
    &=\frac{(2\pi i)^3}{4}\sum_{n=0}^{\infty}&&\binom{2n}{n}^4\left(\frac{\phi}{2^8}\right)^{2n}
\end{alignat}

Both of the power series inside the above integral converge for $\frac{\phi}{4}<|s|<4$ and we find that
\begin{equation}
    \label{eq: holomorphic period as a Hadamard integral for pullback of AESZ3}
    \frac{1}{4}\left(1,0,0,0\right)\Sigma_4\Pi^*(\phi)=\oint_{\ell_{-\frac{\phi}{4}}\ell_0\ell_{\frac{\phi}{4}}}\omega^*_{\begin{psmallmatrix} 0 \\ 1\\ \end{psmallmatrix}}(s)\omega^*_{\begin{psmallmatrix} 0 \\ 1\\ \end{psmallmatrix}}\left(\frac{\phi}{s}\right)\frac{ds}{s},
\end{equation}
which should be interpreted as the statement that the union of $\left(0,1\right)^T\otimes\left(0,1\right)^T$ over a loop in the homotopy class $\ell_{-\frac{\phi}{4}}\ell_0\ell_\frac{\phi}{4}$ determines a $3$-cycle in the homology class represented by
\begin{equation}
    \left(\frac{1}{4},0,0,0\right)^T.
\end{equation}
Note that this is a rational homology class, rather than an integral one. This is to be expected, because the map \eqref{eq: map from mirror of four quadrics in P^7 to fibred product of elliptic curves} from the mirror of four quadrics in $\mathbb{CP}^7$ to the fibred product of elliptic surfaces is a degree $8$ map. We, therefore, expect coefficients in $\frac{1}{8}\mathbb{Z}$.

\begin{figure}
    % [trim={left bottom right top},clip]
    \begin{tikzpicture}
    \node[anchor=south west,inner sep=0] (image) at (0,0) {\frame{\includegraphics[width=\textwidth,trim={2cm 22cm 1cm 1cm},clip]{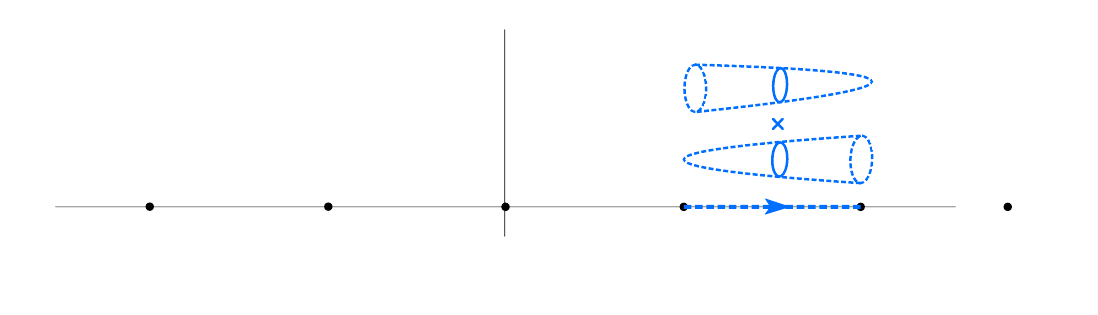}}};
    \begin{scope}[
        x={(image.south east)},
        y={(image.north west)}
    ]
        %\node [white, font=\bfseries] at (0.25,0.65) {$P(X|Y)$};
        %\node at (0.145,0.175) {$s=-4$};
        \node at (0.145,0.270) {$s=-4$};
        \node at (0.14,0.12) {$\begin{psmallmatrix} -1 \\ 2 \end{psmallmatrix}$};
        \node at (0.315,0.425) {$\color{red} s=-\frac{\phi}{4}$};
        \node at (0.30,0.53) {$\color{red} \begin{psmallmatrix} 1 \\ 2 \end{psmallmatrix}$};
        \node[fill=white] at (0.445,0.270) {$s=0$};
        \node at (0.435,0.12) {$\begin{psmallmatrix} 0 \\ 1 \end{psmallmatrix}$};
        \node[fill=white] at (0.445,0.425) {$\color{red} s=0$};
        \node at (0.435,0.53) {$\color{red} \begin{psmallmatrix} 1 \\ 1 \end{psmallmatrix}$};
        \node at (0.585,0.425) {$\color{red} s=\frac{\phi}{4}$};
        \node at (0.585,0.53) {$\color{red} \begin{psmallmatrix} 1 \\ 0 \end{psmallmatrix}$};
        \node at (0.785,0.270) {$s=4$};
        \node at (0.785,0.12) {$\begin{psmallmatrix} 1 \\ 0 \end{psmallmatrix}$};
        \node at (0.925,0.270) {$s=\infty$};
        \node at (0.925,0.12) {$\begin{psmallmatrix} -1 \\ 1 \end{psmallmatrix}$};
        \node at (0.925,0.425) {$\color{red} s=\infty$};
        \node at (0.925,0.53) {$\color{red} \begin{psmallmatrix} 0 \\ 1 \end{psmallmatrix}$};
    \end{scope}
    \end{tikzpicture}
    \captionsetup{format=hang}
    \caption{The complex $s$-plane and singularities for real $\phi$ such that $0<\phi<4^2$. We also visualise a $3$-cycle that vanishes at $\phi=4^2$. Note that at either end of the interval $\frac{\phi}{4}\leq s\leq 4$, one of the circles in the fibre shrinks to a point. This is why the resulting $3$-chain has no $2$-boundary. Points where the first elliptic fibre degenerates have a label below the real axis (in black) and points where the second elliptic fibre degenerates have a label above the real axis (in red). Next to every such point, we list a vanishing cycle, which we compute from the monodromy matrices. Note that we may always approach the same singularity from a different branch. When we do this, the associated vanishing cycle should be multiplied by an appropriate product of the monodromy matrices in Table~\ref{table: monodromy of L_A for omegaxomega}.}
    \label{fig: singularities and a vanishing S^3 on the s-plane for pullback of AESZ 3}
\end{figure}
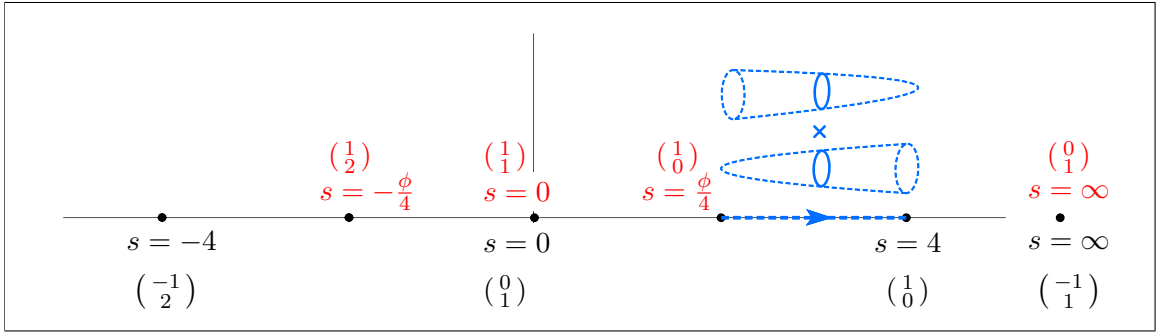

As another example, consider the union of $\left(2,0\right)^T\otimes\left(2,0\right)^T$ over the real interval $\frac{\phi}{4}\leq s\leq 4$. We expect that this is a $3$-cycle, because the class in the fibre is a product of vanishing cycles, with one of them shrinking to zero volume at either end. A useful picture to keep in mind is Figure~\ref{fig: singularities and a vanishing S^3 on the s-plane for pullback of AESZ 3}. This expectation is confirmed by the identity
\begin{equation}
    \label{eq: S^3 Hadamard integral at phi=2^4 for pullback of AESZ3}
    \frac{1}{8}\left(0,0,1,0\right)\Sigma_4\Pi^*(\phi)=\int_{\frac{\phi}{4}}^{4}\omega^*_{\begin{psmallmatrix} 2 \\ 0\\ \end{psmallmatrix}}(s)\omega^*_{\begin{psmallmatrix} 2 \\ 0\\ \end{psmallmatrix}}\left(\frac{\phi}{s}\right)\frac{ds}{s},
\end{equation}
where the above contour is taken to be a straight line in the $s$-plane that starts at $\frac{\phi}{4}$ and ends at $4$. This identity may be proven by noting that the periods in the integrand are holomorphic at $s=4$ (as expected of an integral over a vanishing cycle). We may expand all terms inside the integral around $(s-4)$ and evaluate the integral term-by-term in powers of $\left(\phi-4^2\right)$ (see the computation of \eqref{eq: Hadamard integral with boundary for pullback of AESZ 3}). This just leaves the problem of analytically continuing the expansion of $\omega^*$ around $s=4$ to $s=0$ and the expansion of $\Pi^*$ around $\phi=4^2$ to $\phi=0$. Both $\omega^*$ and $\Pi^*$ are the solutions of the pull-back of hypergeometric differential operators, so the computation may be carried out by use of Mellin-Barnes integrals. This approach, while rigorous, is rather cumbersome when one wants to find many identities like \eqref{eq: S^3 Hadamard integral at phi=2^4 for pullback of AESZ3} . In practice, a much more effective approach is to numerically evaluate $\Pi^*(\phi)$ and the right hand side of \eqref{eq: S^3 Hadamard integral at phi=2^4 for pullback of AESZ3} at some fixed $\phi$. We expect that the components of $\Pi^*(\phi)$ are independent transcendental numbers for generic $\phi$, so we may use a computer to search for an integer relation between the numerical components of $\Pi^*(\phi)$ and the integral on the right hand side of \eqref{eq: S^3 Hadamard integral at phi=2^4 for pullback of AESZ3}. Such an integer relation may be quickly found with the Lenstra–Lenstra–Lovász (LLL) lattice basis reduction algorithm.\cite{lenstra1982factoring}
Note that the union of $\left(2,0\right)^T\otimes\left(2,0\right)^T$ over the real interval $\frac{\phi}{4}\leq s\leq 4$ is a vanishing $3$-cycle, because the two ends of the interval on the $s$-plane approach each other as $\phi\rightarrow 4^2$. This expectation is confirmed by the fact that $(0,0,1,0)$ is the vanishing cycle at $\phi=4^2$, which follows from the monodromy matrix $M_{\phi=4^2}$ in \eqref{eq: monodromy matrices for pullback of AESZ3}, which takes the Picard-Lefschetz form
\begin{equation}
    M_{\phi=4^2}=\mathds{1}_2-
    \begin{pmatrix}
    ~0~\\
    0\\
    1\\
    0
    \end{pmatrix}\begin{pmatrix}0, & 0, & 1, & 0\end{pmatrix}\Sigma_4^T
\end{equation}
The reader should also note that we have used the fibre $(2,0)^T\otimes(2,0)^T$ instead of $(1,0)^T\otimes(1,0)^T$ in the integrand in \eqref{eq: S^3 Hadamard integral at phi=2^4 for pullback of AESZ3}. This is because we only expect a denominator as large as $8$ on the left hand side of \eqref{eq: S^3 Hadamard integral at phi=2^4 for pullback of AESZ3}, and $(1,0)^T\otimes(1,0)^T$ would give us a denominator of $32$. We note the extra factors of $2$ agree with $c_{4}=-2$ and $\widetilde{c}_{\frac{\phi}{4}}=-2$ that we compute from the monodromy matrices in Table~\ref{table: monodromy of L_A for omegaxomega}. A precise explanation for these extra factors of $2$ is likely to come from a careful study of the map \eqref{eq: map from mirror of four quadrics in P^7 to fibred product of elliptic curves} from the mirror of the intersection of four quadrics in $\mathbb{CP}^7$ to the fibred product of two families of elliptic curves. We leave this problem for future investigations.

An identity analogous to \eqref{eq: S^3 Hadamard integral at phi=2^4 for pullback of AESZ3} for the vanishing $3$-cycle at $\phi=-4^2$ is given by
\begin{equation}
    \frac{1}{8}\left(-8,8,-1,-1\right)\Sigma_4\Pi^*(\phi)=\int_{\frac{\phi}{4}}^{-4}\omega^*_{\begin{psmallmatrix} -2 \\ 4\\ \end{psmallmatrix}}(s)\omega^*_{\begin{psmallmatrix} 2 \\ 0\\ \end{psmallmatrix}}\left(\frac{\phi}{s}\right)\frac{ds}{s},
\end{equation}
where the above contour is taken along the upper half $s$-plane for real $\phi$. Note that the $1$-cycles in the integrand match the vanishing cycles at either end of the contour in the $s$-plane. As expected, the union of $\left(-2,4\right)^T\otimes\left(2,0\right)^T$ over a contour between $-4$ and $\frac{\phi}{4}$ on the upper half $s$-plane determines a vanishing cycle at $\phi=-4^2$. We confirm this by noting that the monodromy matrix $M_{\phi=-4^2}$ in \eqref{eq: monodromy matrices for pullback of AESZ3} has the Picard-Lefschetz form
\begin{equation}
    M_{\phi=-4^2}=\mathds{1}_2-
    \begin{pmatrix}
    -8~\\
    8\\
    -1\\
    -1
    \end{pmatrix}\begin{pmatrix}-8, & 8, & -1, & -1\end{pmatrix}\Sigma_4^T~.
\end{equation}

We consider yet another example in order to illustrate that, in general, one must really consider collections of contours on the $s$-plane, each with its own choice of homology class in the fibre. We numerically confirm that
\begin{align}
\label{eq: contour on positive real s-line for pullback of AESZ3}
    \begin{split}
        \frac{1}{8}\left(0,-1,0,0\right)\Sigma_4\Pi^*(\phi)=&~\int_{0}^{\frac{\phi}{4}}\omega^*_{\begin{psmallmatrix} 0 \\ 1 \end{psmallmatrix}}(s)\omega^*_{\begin{psmallmatrix} 1 \\ 1 \end{psmallmatrix}}\left(\frac{\phi}{s}\right)\frac{ds}{s}\\
        &+\int_{\frac{\phi}{4}}^{4}\omega^*_{\begin{psmallmatrix} 0 \\ 1 \end{psmallmatrix}}(s)\omega^*_{\begin{psmallmatrix} 0 \\ 1 \end{psmallmatrix}}\left(\frac{\phi}{s}\right)\frac{ds}{s}\\
        &+\int_{4}^{\infty}\omega^*_{\begin{psmallmatrix} -1 \\ ~1 \end{psmallmatrix}}(s)\omega^*_{\begin{psmallmatrix} 0 \\ 1 \end{psmallmatrix}}\left(\frac{\phi}{s}\right)\frac{ds}{s}.
    \end{split}
\end{align}
For example, at the regular point $\phi=8$, both the left and right hand side of \eqref{eq: contour on positive real s-line for pullback of AESZ3} evaluate to
\begin{equation}
    \frac{1}{8}\left(0,-1,0,0\right)\Sigma_4\Pi^*(8)=-34.44699775873907269740146097820\ldots~.
\end{equation}
Note that each of the above integrals are integrals over a $3$-chain with boundary, but the sum defines a $3$-cycle. For example, $(0,1)^T\otimes(1,1)^T$ does not match the vanishing cycle at $s=\frac{\phi}{4}$ (see Table~\ref{table: monodromy of L_A for omegaxomega} and Figure~\ref{fig: singularities and a vanishing S^3 on the s-plane for pullback of AESZ 3}). However, the boundary term is cancelled by the next integral because
\begin{equation}
    \begin{pmatrix} 1\\1\end{pmatrix}\equiv\begin{pmatrix} 0\\ 1\end{pmatrix}~\text{mod}~\widetilde{v}_{\frac{\phi}{4}}~.
\end{equation}
Geometrically, one should imagine gluing two circles together, one in the homology class $(1,1)^T$ and the other in the homology class $(1,0)^T$. This is only possible at a point where the elliptic curve degenerates and the two $1$-cycles are equal modulo the vanishing cycle. The same kind of cancellation takes place between the second and third integrals i.e.
\begin{equation}
    \begin{pmatrix} 0 \\ 1\end{pmatrix}\equiv\begin{pmatrix} -1\\1\end{pmatrix}~\text{mod}~v_{4}~.
\end{equation}
 This illustrates the point that the most general identity of the form \eqref{eq: integral over a network equals a period} involves a collection of contours with different choices of homology classes in the fibre.

\subsection{Extensions of AESZ 3 and its double cover}
\label{subsec: Extensions of AESZ 3 and its double cover}
We have seen that, for judicious choices of contours $\ell^{(i)}$ between two singularities in the $s$-plane and homology classes $\gamma_1^{(i)}$ and $\gamma_2^{(i)}$ in the fibres, the integral \eqref{eq: integral over a network equals a period} is a solution of a Picard-Fuchs equation.\footnote{The reader might argue that a closed contour in the $s$-plane need not start and end at a singularity. This is the case, for example, in \eqref{eq: holomorphic period as a Hadamard integral for pullback of AESZ3}. However, we may always deform such a contour so that it starts and ends at some singularity without changing the associated integral. Moreover, we will assume that this can be done even when the contours are constrained by the special Lagrangian condition in Section~\ref{sec: BPS networks from special Lagrangian condition}.} However, generic choices of contours and fibres will lead to a $3$-chain with boundaries at the singularities in the $s$-plane. In other words,
\begin{equation}
\mathcal{L}_{\text{AESZ3}}^*\left(\sum_{i=1}^{N}\int_{\ell^{(i)}}\omega^*_{\gamma^{(i)}_1}(s)\omega^*_{\gamma^{(i)}_2}\left(\frac{\phi}{s}\right)\frac{ds}{s}\right)\neq0~.
\end{equation}
In the case of AESZ 3 and its pullback, we have been able to find a simple extension of the differential equation that annihilates the above sum of integrals. The extension is given by
\begin{equation}
    \label{eq: extension of pullback of AESZ 3}
    \bigg((\phi^2-2^8)\theta_\phi^2+2(\phi^2+2^9)\theta_\phi+(\phi^2-2^{10})\bigg)\mathcal{L}_{\text{AESZ3}}^*,
\end{equation}
and it annihilates the integral
\begin{equation}
    \label{eq: generic Hadamard integral for pullback of AESZ 3}
    \int_{\ell}\omega^*_{\gamma_1}(s)\omega^*_{\gamma_2}\left(\frac{\phi}{s}\right)\frac{ds}{s}
\end{equation}
for all possible choices of contours $\ell$ between two singularities and fibre classes $\gamma_1$ and $\gamma_2$. The extended Picard-Fuchs operator has the Riemann symbol
\begin{equation}
    \mathcal{P}\left\{\begin{array}{c c c} 0 & \pm 4^2 & \infty\\
    \hline
    0 & 0 & 1\\
    0 & 1 & 1\\
    0 & 1 & 1\\
    0 & 2 & 1\\
    2 & 2 & 3\\
    2 & 2 & 3\\    
    \end{array}~; \phi\right\}
\end{equation}
and is also invariant under the $\phi\rightarrow\frac{2^8}{\phi}$ symmetry of $\mathcal{L}_\text{AESZ3}^*$.

The above extension of $\mathcal{L}_\text{AESZ3}^*$ may be found as follows. First, note that $\omega^*(s)$ has the following expansion around $s=4$
\begin{equation}
    \omega^*(s)=\begin{pmatrix} \left(\pi-5i\log 2\right) & i\\ -\pi & 0\end{pmatrix}\left(\begin{array}{l}f_0^{(4)}(s) \\ f_0^{(4)}(s)\log\left(s-4\right)+f_1^{(4)}(s)\end{array}\right)
\end{equation}
where
\begin{align}
    \begin{split}
        f_0^{(4)}(s)&=\sum_{n=0}^\infty a_n^{(4)} (s-4)^n\\
        &=1-\frac{1}{8}(s-4)+\frac{5}{256}(s-4)^2-\frac{7}{2048}(s-4)^3+\frac{169}{262144}(s-4)^4+\ldots\\
        f_1^{(4)}(s)&=\sum_{n=0}^\infty b_n^{(4)} (s-4)^n\\
        &=-\frac{1}{8}(s-4)+\frac{7}{256}(s-4)^2-\frac{17}{3072}(s-4)^3+\frac{1775}{1572864}(s-4)^4+\ldots~.
    \end{split}
\end{align}
The coefficients $a_n^{(4)}$ and $b_n^{(4)}$ may be computed recursively with the Picard-Fuchs equation. We have computed the above transition matrix numerically to a high precision and identified the terms appearing the matrix. Alternatively, one may use the fact that $\omega^*(s)$ is annihilated by the pull-back of a hypergeometric differential operator, so the transition matrices may be computed with the help of Mellin-Barnes integrals. Now consider the integral
\begin{equation}
   \int_{\frac{\phi}{4}}^{4}\omega^*_{\begin{psmallmatrix} 1 \\ 0 \end{psmallmatrix}}(s)\omega^*_{\begin{psmallmatrix} 1 \\ 0 \end{psmallmatrix}}\left(\frac{\phi}{s}\right)\frac{ds}{s}=\pi^2\int_{\frac{\phi}{4}}^4 f_0^{(4)}(s)f^{(4)}_0\left(\frac{\phi}{s}\right)\frac{ds}{s}
\end{equation}
where $0<\phi<4^2$ and the contour between $s=\frac{\phi}{4}$ and $s=4$ is taken to be a straight line. We saw in \eqref{eq: S^3 Hadamard integral at phi=2^4 for pullback of AESZ3} that this determines a solution of $\mathcal{L}^*_{\text{AESZ3}}$. With this choice of contour, the only way of matching the vanishing cycles at $s=\frac{\phi}{4}$ and $s=4$ is with a choice of fibre class proportional to $(1,0)^T\otimes(1,0)^T$, as in \eqref{eq: S^3 Hadamard integral at phi=2^4 for pullback of AESZ3}. It follows that
\begin{equation}
    \label{eq: Hadamard integral with boundary for pullback of AESZ 3}
    I(\phi)=\int_{\frac{\phi}{4}}^4\left(f_0^{(4)}(s)\log(s-4)+f_1^{(4)}(s)\right)f^{(4)}_0\left(\frac{\phi}{s}\right)\frac{ds}{s}
\end{equation}
cannot be proportional to an integral over a $3$-cycle, so we expand $I(\phi)$ around the singular point $\phi=4^2$ and look for an extension of $\mathcal{L}^*_{\text{AESZ3}}$ that annihilates it. We expand the $f_j^{(4)}$ as power series, swap the order of summation and integration and replace each occurrence of $s$ with $(s-4)$ e.g.
\begin{equation}
    \left(\frac{\phi}{s}-4\right)^n=\sum_{k=0}^n\binom{n}{k}\frac{4^{n-k}}{s^n}(-1)^{n-k}(s-4)^{n-k}(\phi-4^2)^k
\end{equation}
and
\begin{equation}
    \frac{1}{s^{n+1}}=\sum_{l=0}^\infty\binom{n+l}{l}\frac{(-1)^l}{4^{n+l+1}}(s-4)^l.
\end{equation}
We then use the identity
\begin{equation}
    \int x^n \log x~dx=-\frac{x^{n+1}}{(n+1)^2}+\frac{x^{n+1}}{n+1}\log x
\end{equation}
and find that
\begin{equation}
    I(\phi)=A(\phi) \log(\phi-4^2)-\log(4)A(\phi)+B(\phi)+C(\phi)
\end{equation}
where
\begin{alignat}{2}
        A(\phi)&=\sum_{n=0}^\infty\Bigg(\sum_{i=0}^n\sum_{j=0}^{n-i}&&(-1)^{n-i-j+1}\frac{a_i^{(4)}a_j^{(4)}}{4^{2n+2-i-j}}\frac{(n-i)!(n-j)!}{(n-i-j)!(n+1)!}\Bigg)(\phi-4^2)^{n+1}\\
        B(\phi)&=\sum_{n=0}^\infty\Bigg(\sum_{i=0}^n\sum_{j=0}^{n-i}&&(-1)^{n-i-j}\frac{a_i^{(4)}a_j^{(4)}}{4^{2n+2-i-j}}\frac{(n-i)!(n-j)!}{(n-i-j)!(n+1)!}~\\
        \nonumber & ~&&\times\left(H_{n+1}-H_{n-j}\right)\Bigg)(\phi-4^2)^{n+1}\\
        C(\phi)&=\sum_{n=0}^\infty\Bigg(\sum_{i=0}^n\sum_{j=0}^{n-i}&&(-1)^{n-i-j+1}\frac{a_i^{(4)}b_j^{(4)}}{4^{2n+2-i-j}}\frac{(n-i)!(n-j)!}{(n-i-j)!(n+1)!}\Bigg)(\phi-4^2)^{n+1}.
\end{alignat}
and $H_n$ is the harmonic sum
\begin{equation}
    H_n=\sum_{k=1}^n\frac{1}{k}.
\end{equation}
In deriving these formulae, we have used the identities
\begin{align}
    \begin{split}
        \sum_{k=0}^n\binom{n}{k}\frac{(-1)^k}{m+n+l+1-k}&=\frac{(-1)^n(m+l)!n!}{(m+n+l+1)!}\\
        \sum_{k=0}^n\binom{n}{k}\frac{(-1)^k}{(m+n+l+1-k)^2}&=\frac{(-1)^n(m+l)!n!}{(m+n+l+1)!}\left(H_{m+n+l+1}-H_{m+l}\right)~.
    \end{split}
\end{align}
It is now straightforward, with the aid of a computer, to expand $I(\phi)$ to some finite order in $(\phi-4^2)$ and search for an extension of $\mathcal{L}^*_{\text{AESZ3}}$ that annihilates it. In this way, we find \eqref{eq: extension of pullback of AESZ 3}. For various real values of $\phi$ satisfying $0<|\phi|<2^4$, we numerically confirm that the extension annihilates all integrals of the form \eqref{eq: generic Hadamard integral for pullback of AESZ 3} where $\ell$ is some contour between the two singularities and
\begin{equation}
    \gamma_1\otimes\gamma_2=e_i\otimes e_j~~~~~~\text{where}~~~~~~e_1=\begin{pmatrix}1\\0\end{pmatrix}~~~~~~\text{and}~~~~~~e_2=\begin{pmatrix}0\\1\end{pmatrix}.
\end{equation}
% \begin{equation}
%     \gamma_1\otimes\gamma_2\in\left\{
%     \begin{pmatrix}1 \\ 0\end{pmatrix}\otimes \begin{pmatrix}1 \\ 0\end{pmatrix},
%     \begin{pmatrix}1 \\0\end{pmatrix}\otimes \begin{pmatrix}0 \\ 1\end{pmatrix},
%     \begin{pmatrix}0 \\ 1\end{pmatrix}\otimes \begin{pmatrix}1 \\ 0\end{pmatrix},
%     \begin{pmatrix}0 \\ 1\end{pmatrix}\otimes \begin{pmatrix}0 \\ 1\end{pmatrix}\right\}~.
% \end{equation}
All integrals of the form \eqref{eq: generic Hadamard integral for pullback of AESZ 3} can be written as a linear combination of such integrals, so there are only finitely many integrals to check for each $\phi$. We conjecture that this is true for all $\phi$.

Recall that AESZ 3 and its pull-back are related by $\varphi=\left(\frac{\phi}{2^8}\right)^2$. A similar exercise for $\mathcal{L}_{\text{AESZ3}}$ reveals that
\begin{equation}
    \bigg((2^8\varphi^2-1)\theta_\varphi^2+2(2^7\varphi^2+1)\theta_\varphi+(2^6\varphi^2-1)\bigg)\mathcal{L}_{\text{AESZ3}}
\end{equation}
annihilates the corresponding integrals on the $s$-plane. It follows that the extension of $\mathcal{L}^*_{\text{AESZ3}}$ is a pull-back of the extension of $\mathcal{L}_{\text{AESZ3}}$.

\subsection{AESZ 101}

\begin{figure}
    % [trim={left bottom right top},clip]
    \begin{tikzpicture}
    \node[anchor=south west,inner sep=0] (image) at (0,0) {\frame{\includegraphics[width=\textwidth,trim={0 19cm 0 1cm},clip]{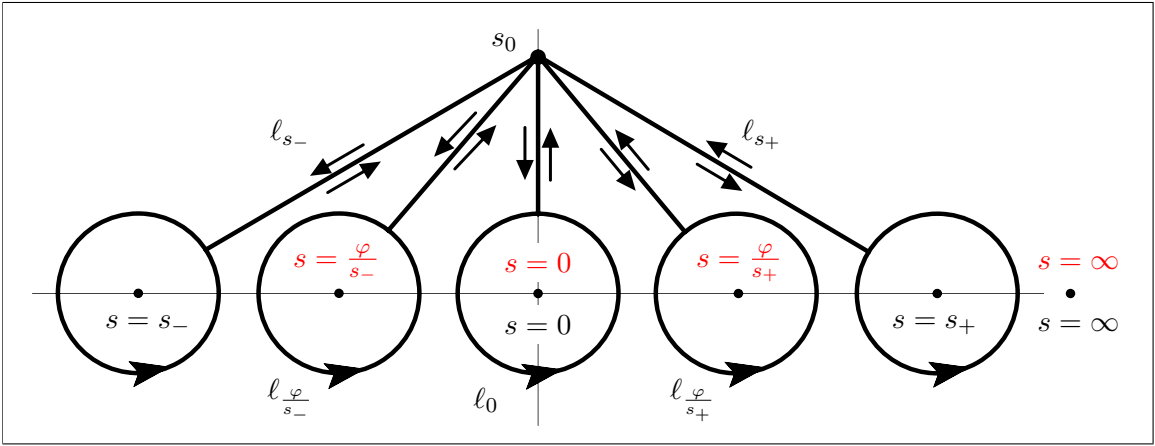}}};
    \begin{scope}[
        x={(image.south east)},
        y={(image.north west)}
    ]
        %\node [white, font=\bfseries] at (0.25,0.65) {$P(X|Y)$};
        \node at (0.126,0.27) {$s=s_-$};
        \node at (0.29,0.41) {$\color{red} s=\frac{\varphi}{s_-}$};
        \node[fill=white] at (0.465,0.27) {$s=0$};
        \node[fill=white] at (0.465,0.41) {$\color{red} s=0$};
        \node at (0.64,0.41) {$\color{red} s=\frac{\varphi}{s_+}$};
        \node at (0.81,0.27) {$s=s_+$};
        \node at (0.935,0.27) {$s=\infty$};
        \node at (0.935,0.41) {$\color{red} s=\infty$};
        \node at (0.25,0.70) {$\ell_{s_-}$};
        \node at (0.25,0.10) {$\ell_{\frac{\varphi}{s_-}}$};
        \node at (0.42,0.10) {$\ell_{0}$};
        \node at (0.60,0.10) {$\ell_{\frac{\varphi}{s_+}}$};
        \node at (0.66,0.70) {$\ell_{s_+}$};
        \node at (0.436,0.91) {$s_0$};
    \end{scope}
    \end{tikzpicture}
    \captionsetup{format=hang}
    \caption{Generators of $\pi_1\left(\mathbb{CP}^1\backslash\{s_-,\frac{\varphi}{s_-},0,\frac{\varphi}{s_+},s_+,\infty\},s_0\right)$ in the complex $s$-plane for real $\varphi$ such that $0<\varphi<s_+^2$. We have listed points on the $s$-plane where the first elliptic curve degenerates below the real axis (in black) and points where the second elliptic curve degenerates above the real axis (in red). Note that we have chosen a basepoint $s_0$ with very large imaginary part.}
    \label{fig: generators of fundamental group of s-plane for AESZ101}
\end{figure}

In this section, we compute solutions of AESZ 101 by computing integrals over $3$-cycles on the fibred product of elliptic surfaces. The analysis is very similar to that of AESZ 3, so this section will be rather terse. Let $s_\pm=\frac{1}{2}\left(-11\pm5\sqrt 5\right)$. We must fix the branches that $\omega(s)$ and $\omega\left(\frac{\varphi}{s}\right)$ start on. We do this by choosing some real $0<\varphi<s_+^2$ and requiring that $\omega(s)$ and $\omega\left(\frac{\varphi}{s}\right)$ are on the same branch when evaluated on the interval $\frac{\varphi}{s_+}<s<s_+^2$. This defines $\omega(s)\otimes\omega\left(\frac{\varphi}{s}\right)$ by analytic continuation. We choose generators of the fundamental group as in Figure~\ref{fig: generators of fundamental group of s-plane for AESZ101} and find the monodromy matrices in Table~\ref{table: monodromy of L_b for omegaxomega}. From this, we compute the vanishing cycles at the various singularities in the $s$-plane, which we also list in Table~\ref{table: monodromy of L_b for omegaxomega}.

\begin{table}
\centering
\begin{tabular}{|c||c|c|c|c|c|c|}
\hline
$s_*$ & $m_{s_*}$ & $c_{s_*}$ & $v_{s_*}$ & $\widetilde{m}_{s_*}$ & $\widetilde{c}_{s_*}$ & $\widetilde{v}_{s_*}$ \\
\hline
\hline
&&&&&&\\[-9pt]
%\hhline{=======}
$s_-=\frac{1}{2}(-11- 5\sqrt{5})$ & $\left(\begin{matrix} 11 & -4 \\ 25 & -9 \end{matrix}\right)$ & $1$ & $\begin{pmatrix} 2 \\ 5\end{pmatrix}$ & $\mathds{1}_2$ & n/a & n/a \\[15pt]
\hline
&&&&&&\\[-9pt]
$\frac{\varphi}{s_-}$ & $\mathds{1}_2$ & n/a & n/a & $\begin{pmatrix} -9 & -4 \\ 25 & 11\end{pmatrix}$ & 1 & $\begin{pmatrix} -2 \\ 5\end{pmatrix}$ \\[15pt]
\hline
&&&&&&\\[-9pt]
0 & $\begin{pmatrix} 1 & 0\\ 5 & 1\end{pmatrix}$ & $5$ & $\begin{pmatrix} 0 \\1\end{pmatrix}$ & $\begin{pmatrix} -9 & -5\\ 20 & 11\end{pmatrix}$ & $5$ & $\begin{pmatrix}-1 \\2\end{pmatrix}$ \\[15pt]
\hline
&&&&&&\\[-9pt]
$s_+=\frac{1}{2}(-11+ 5\sqrt{5})$ & $\begin{pmatrix} 1 & -1\\ 0 & 1\end{pmatrix}$ & $1$ & $\begin{pmatrix} 1 \\ 0\end{pmatrix}$ & $\mathds{1}_2$ & n/a & n/a \\[15pt]
\hline
&&&&&&\\[-9pt]
$\frac{\varphi}{s_+}$ & $\mathds{1}_2$ & n/a & n/a & $\begin{pmatrix} 1 & -1\\ 0 & 1\end{pmatrix}$ & $1$ & $\begin{pmatrix} 1 \\ 0\end{pmatrix}$ \\[15pt]
\hline
&&&&&&\\[-9pt]
$\infty$ & $\begin{pmatrix} 11 & -5\\ 20 & -9\end{pmatrix}$ & $5$ & $\begin{pmatrix} 1 \\ 2\end{pmatrix}$ & $\begin{pmatrix} 1 & 0 \\ 5 & 1\end{pmatrix}$ & $5$ & $\begin{pmatrix}0 \\ 1\end{pmatrix}$ \\[15pt]
\hline
\end{tabular}
\captionsetup{format=hang}
\caption{The monodromy matrices of the elliptic periods associated with the Hadamard product $\mathcal{L}_{\text{AESZ101}}$.}
\label{table: monodromy of L_b for omegaxomega}
\end{table}

The fibre class $(0,1)^T\otimes (0,1)^T$ is invariant under monodromy around a closed contour in the homotopy class $\ell_{\frac{\varphi}{s_-}}\ell_0\ell_{\frac{\varphi}{s_+}}$. The union of this fibre class over such a contour defines a topological $T^3$. By integrating over this $T^3$, we recover the defining property of a Hadamard product i.e.
\begin{equation}
    \left(1,0,0,0\right)\Sigma_4\Pi(\varphi)=\oint_{\ell_{\frac{\varphi}{s_-}} \ell_0 \ell_{\frac{\varphi}{s_+}}}
    \omega_{\begin{psmallmatrix} 0 \\ 1\\ \end{psmallmatrix}}(s)
    \omega_{\begin{psmallmatrix} 0 \\ 1\\ \end{psmallmatrix}}\left(\frac{\varphi}{s}\right)
    \frac{ds}{s}.
    \label{eq: identity for Hadamard integral associated to vanishing cycle of AESZ 101 at nearest conifold}
\end{equation}
Note that, unlike in \eqref{eq: holomorphic period as a Hadamard integral for pullback of AESZ3} there are no factors of $\frac{1}{4}$ in this expression. Such factors in \eqref{eq: holomorphic period as a Hadamard integral for pullback of AESZ3} and related expression arise from the fact that the map from the smooth Calabi-Yau threefold (which determines the basis of periods $\Pi^*$) to the fibred product of elliptic surfaces is a degree $8$ map. For the geometries associated to AESZ 101, the analogous map is birational.

Let $\varphi$ be real and $0<\varphi<s_+^2$. The singularities on the $s$-plane are all real and we may identify the vanishing cycle at the singularity $\varphi=s_+^2$ with the union of $(1,0)^T\otimes(1,0)^T$ over the real interval $\frac{\varphi}{s_+}\leq s\leq s_+$. This is confirmed by the identity
\begin{equation}
    \label{eq: vanishing S3 Hadamard integral at nearest conifold of AESZ 101}
    \left(0,0,1,0\right)\Sigma_4\Pi(\varphi)=\int_\frac{\varphi}{s_+}^{s_+}\omega_{\begin{psmallmatrix} 1 \\ 0\\ \end{psmallmatrix}}(s)
    \omega_{\begin{psmallmatrix} 1 \\ 0\\ \end{psmallmatrix}}\left(\frac{\varphi}{s}\right)
    \frac{ds}{s},
\end{equation}
which may be checked numerically. Alternatively, one may evaluate the integral on the right hand side as a power series around $\varphi=s_+^2$, along the lines of the calculation in subsection~\ref{subsec: Extensions of AESZ 3 and its double cover}.

As a final example, we find that
\begin{align}
    \begin{split}
        \left(0,-4,0,0\right)\Sigma_4\Pi(\varphi)=&~\int_{0}^{\frac{\varphi}{s_+}}\omega_{\begin{psmallmatrix} 0 \\ 2 \end{psmallmatrix}}(s)\omega_{\begin{psmallmatrix} -1 \\ 2 \end{psmallmatrix}}\left(\frac{\varphi}{s}\right)\frac{ds}{s}\\
        &+\int_{\frac{\varphi}{s_+}}^{s_+}\omega_{\begin{psmallmatrix} 0 \\ 2 \end{psmallmatrix}}(s)\omega_{\begin{psmallmatrix} 0 \\ 2 \end{psmallmatrix}}\left(\frac{\varphi}{s}\right)\frac{ds}{s}\\
        &+\int_{s_+}^{\infty}\omega_{\begin{psmallmatrix} 1 \\ 2 \end{psmallmatrix}}(s)\omega_{\begin{psmallmatrix} 0 \\ 2 \end{psmallmatrix}}\left(\frac{\varphi}{s}\right)\frac{ds}{s}.
    \end{split}
\end{align}
This identity may be confirmed numerically for real $0\leq \varphi\leq s_+^2$, where the each of the contours on the $s$-plane are taken along the positive real axis. As in the case of \eqref{eq: contour on positive real s-line for pullback of AESZ3}, none of the above integrals are themselves annihilated by $\mathcal{L}_{\text{AESZ101}}$. However, all boundary terms cancel in the sum, which is annihilated by $\mathcal{L}_{\text{AESZ101}}$.

%---------------------------------------------------------------------------------------------------------------
\section{BPS networks}
\label{sec: BPS networks from special Lagrangian condition}
\subsection{Networks from special Lagrangian condition}
A D3-brane on a Calabi-Yau threefold $X$ at zero string coupling is determined by a special Lagrangian submanifold $L$ that supports a flat $U(1)$ connection.\cite{Becker:1995kb} The submanifold $L$ has three real dimensions and is defined by an inclusion map $\iota: L\hookrightarrow X$ such that
\begin{align}
    \label{eq: special Lagrangian condition}
    \begin{split}
        \iota^*\omega_K &=0\\
        \iota^*\text{Im}~e^{-i\theta}\Omega &=0
    \end{split}
\end{align}
where $\omega_K$ and $\Omega$ are the K\"{a}hler form and holomorphic volume form of the Calabi-Yau threefold, respectively.\footnote{ Here we denote the K\"{a}hler form by $\omega_K$ to avoid clashing with $\omega$, which we have used to denote the periods of the holomorphic $1$-form on an elliptic curve.} The angle $\theta$ accounts for the fact that the normalisation of $\Omega$ is arbitrary and, for a given submanifold $L$, it can be removed by rescaling $\Omega$. However, it becomes important when one wants to consider the stability of multiple special Lagrangian submanifolds at the same time. The flat $U(1)$ connections on $L$ are classified by the topology of $L$, so problem of finding stable D3-branes boils down to the problem of solving \eqref{eq: special Lagrangian condition}. Unfortunately, generic solutions of \eqref{eq: special Lagrangian condition} are extremely difficult to come by. Most known solutions are constructed as the fixed loci of an anti-holomorphic involution on $X$. For example, when the polynomials defining $X$ are defined over the real numbers, $X$ is invariant under complex conjugation and the real locus is a union of one or more disconnected special Lagrangian submanifolds.  See \cite{Walcher:2006rs} for an example. 

In previous sections, we have studied the fibred product of two families of elliptic curves with the holomorphic $3$-form
\begin{equation}
    \label{eq: 3form on fibred product of elliptic curves}
    \frac{dx_1}{2y_1(s)}\wedge\frac{dx_2}{2y_2\left(\frac{\varphi}{s}\right)}\wedge\frac{ds}{s},
\end{equation}
where $x_i$ and $y_i$ are the variables used in the Weierstrass equation of the corresponding elliptic curve. We have been able to identify $3$-cycles on such geometries with $3$-cycles on a compact Calabi-Yau threefold $X_\varphi$. One naturally asks - can we also impose the special Lagrangian condition? The K\"{a}hler form depends on the Ricci flat Calabi-Yau metric, which is difficult to compute. However, we have a lot of control over the holomorphic $3$-form. For generic $\varphi$, the integral of the holomorphic $3$-form on a $3$-cycle is non-vanishing, so we must consider $3$-cycles obtained as the union of a circle in each of the elliptic fibres over a contour on $\mathbb{CP}^1$. By integrating \eqref{eq: 3form on fibred product of elliptic curves} over a product of two circles, one in each of the elliptic fibres, we are led to the constraint
\begin{equation}
    \label{eq: constant phase condition}
    \omega_{\gamma_1}(s)\omega_{\gamma_2}\left(\frac{\varphi}{s}\right)\frac{1}{s}\frac{ds}{dl}=e^{i\theta},
\end{equation}
for the contour $s=s(l)$ on the base. We have chosen to use a parameterisation by $l$ such that the absolute value of the right hand side is equal to unity. Here, as in previous sections, $\omega_{\gamma_i}$ is the period of the holomorphic $1$-form on an elliptic fibre over a $1$-cycle in the homology class $\gamma_i$.

Solutions of the constant phase condition \eqref{eq: constant phase condition} define a network of contours on $\mathbb{CP}^1\backslash\Delta_1\cup\Delta_2$, where $\Delta_i$ is the finite set of points on $\mathbb{CP}^1$ where the $i^{\text{th}}$ elliptic curve degenerates. They are strongly reminiscent of the spectral and exponential networks that have previously appeared in string theory literature. This is not surprising, because they all arise in essentially the same way. In the context of the Seiberg-Witten analysis of field theories with $\mathcal{N}=2$ supersymmetry, BPS states are enumerated by certain critical spectral networks.\cite{Klemm:1996bj,Gaiotto:2012rg} These are constant phase contours for a differential on a punctured Riemann surface. In \cite{Klemm:1996bj}, the Seiberg-Witten differential arises from a reduction of a holomorphic $3$-form on a Calabi-Yau threefold to the base of a fibration, which is quite similar in spirit to the construction in this paper. Spectral networks have been generalised to exponential networks, which arise from the special Lagrangian condition on toric Calabi-Yau threefolds.\cite{Eager:2016yxd,Banerjee:2018syt,Banerjee:2020moh,Banerjee:2022oed,Banerjee:2023zqc} 

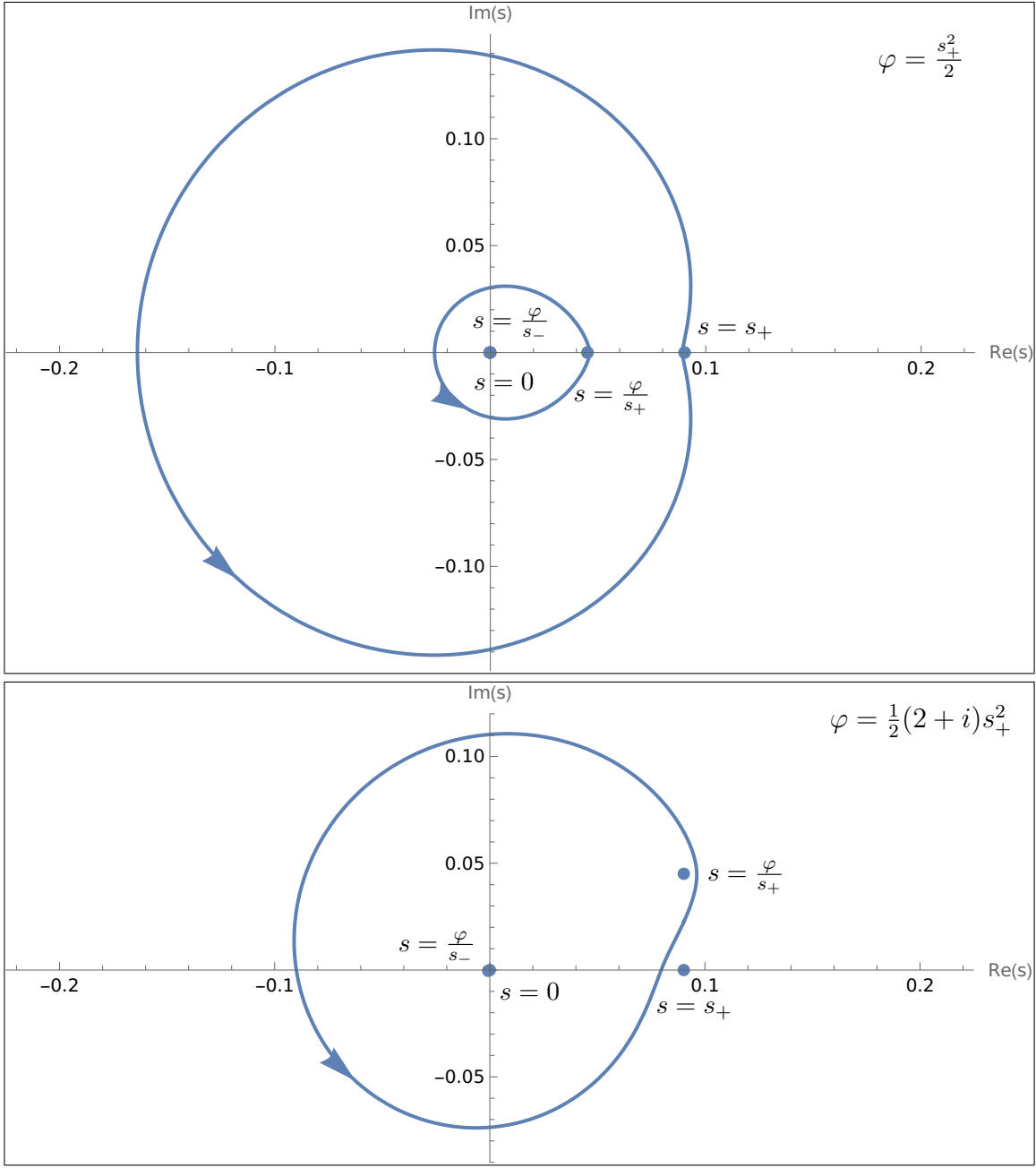
\begin{figure}
    \begin{tikzpicture}
    \node[anchor=south west,inner sep=0] (image) at (0,0) {\frame{\includegraphics[width=\textwidth]{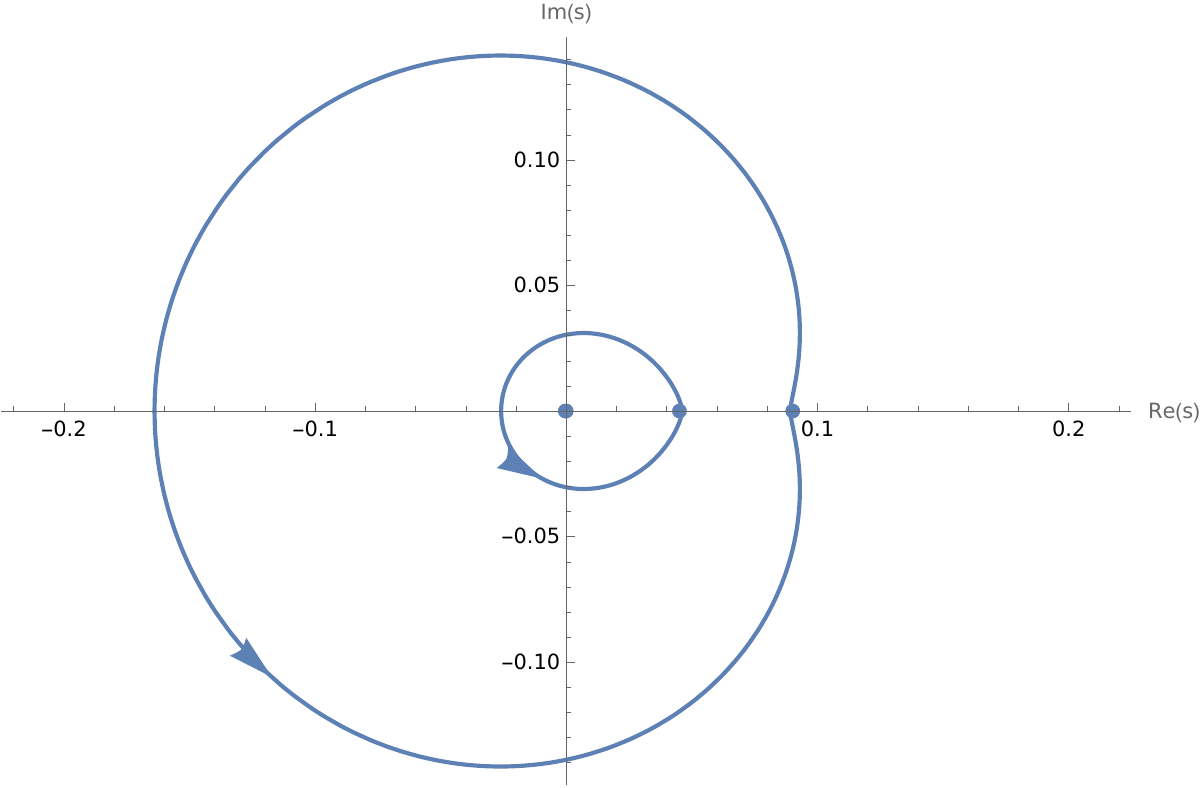}}};
    \begin{scope}[
        x={(image.south east)},
        y={(image.north west)}
    ]
        %\node [white, font=\bfseries] at (0.25,0.65) {$P(X|Y)$};
        \node at (0.89,0.92) {{\large $\varphi=\frac{s_+^2}{2}$}};
        \node at (0.492,0.52) {$s=\frac{\varphi}{s_-}$};
        \node at (0.485,0.435) {$s=0$};
        \node at (0.59,0.41) {$ s=\frac{\varphi}{s_+}$};
        \node at (0.71,0.51) {$s=s_+$};
    \end{scope}
    \end{tikzpicture}

    \vspace{3pt}
        
    \begin{tikzpicture}
    \node[anchor=south west,inner sep=0] (image) at (0,0) {\frame{\includegraphics[width=\textwidth]{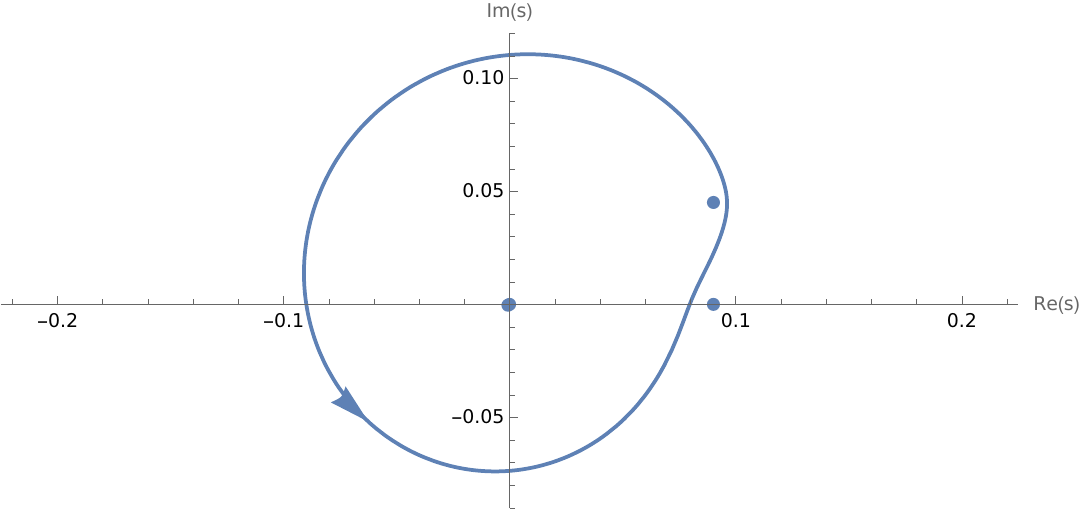}}};
    \begin{scope}[
        x={(image.south east)},
        y={(image.north west)}
    ]
        %\node [white, font=\bfseries] at (0.25,0.65) {$P(X|Y)$};
        \node at (0.89,0.92) {{\large $\varphi=\frac{1}{2}(2+i)s_+^2$}};
        \node at (0.42,0.45) {$s=\frac{\varphi}{s_-}$};
        \node at (0.51,0.36) {$s=0$};
        \node at (0.67,0.32) {$s=s_+$};
        \node at (0.72,0.595) {$s=\frac{\varphi}{s_+}$};
    \end{scope}
    \end{tikzpicture}
    \captionsetup{format=hang}
    \caption{Two plots of constant phase contours with $\varphi=\frac{s_+^2}{2}$ (top plot) and $\varphi=\frac{1}{2}(2+i)s_+^2$ (bottom plot). All three contours in these plots have the associated fibre class $\gamma_1\otimes\gamma_2=(0,1)^T\otimes(0,1)^T$ and they all lift to the homology class $\Gamma=(1,0,0,0)^T$. This determines the associated phases $\theta=\text{arg}\left(\int_\Gamma \Omega(\varphi)\right)=-\frac{\pi}{2}$ for the contours in the top plot and $\theta=\text{arg}\left(\int_\Gamma \Omega(\varphi)\right)\approx-1.49204$ for the contour in the bottom plot. Note that the two contours in the top plot represent the boundaries of a family of constant phase contours that intersect the real line at $\frac{\varphi}{s_+}\leq s\leq s_+$. Also note that the point at the origin represents the two singularities $s=\frac{\varphi}{s_-}$ and $s=0$, which are difficult to distinguish at this scale. The remaining finite singularity at $s=s_-\approx -11.09017$ is not plotted.
    }
    \label{fig: T3 contours at large complex structure}
\end{figure}

To be concrete, we consider AESZ 101. Two examples of solutions of the constant phase condition \eqref{eq: constant phase condition} are shown in the top plot of Figure~\ref{fig: T3 contours at large complex structure}. Both of these contours are topological circles with a monodromy invariant $S^1\times S^1$ as fibres. In other words, they each determine a topological $T^3$. Between these two contours, there is an infinite family of solutions of the constant phase condition \eqref{eq: constant phase condition}. As $\varphi\rightarrow 0$, this family of $T^3$s defines an SYZ like fibration.\cite{Strominger:1996it} Indeed, as computed in \eqref{eq: identity for Hadamard integral associated to vanishing cycle of AESZ 101 at nearest conifold}, the integral over any single $T^3$ is holomorphic at $\varphi=0$, as expected of the mirror of a D0-brane at large volume. In this example, the $T^3$ fibration only reaches up to the singularity $s=s_+$ on the base $\mathbb{CP}^1$.

The fact that the constant phase contours in Figure~\ref{fig: T3 contours at large complex structure} come in families is to be expected. Generally, the embedding of a special Lagrangian submanifold inside a Calabi-Yau threefold may be deformed while preserving the special Lagrangian condition. The resulting moduli space of special Lagrangian submanifolds has real dimension equal to the first Betti number $b_1(L)$. Infinitesimally, such deformations are in one to one correspondence with harmonic $1$-forms on $L$.\cite{1997dg.ga....11002H,Strominger:1996it}

We may also choose $\theta(\varphi)=\text{arg}\left(\int_\Gamma \Omega(\varphi)\right)$ from the beginning, slightly vary the complex structure parameter $\varphi$ and look for solutions of \eqref{eq: constant phase condition} while keeping $\Gamma$ and $\gamma_1\otimes\gamma_2$ fixed. See, for example, the bottom plot of Figure~\ref{fig: T3 contours at large complex structure}. The reader may notice that the constant phase contour appears as if it might collide with the singularity $\frac{\varphi}{s_+}$ for a large enough variation of $\varphi$. This is indeed case and it is expected. Typically, for a small deformation of complex structure, one can preserve the special Lagrangian condition by making a compensating small deformation of the embedding map and the phase $\theta$. However, for a large enough variation, this may no longer be possible and a special Lagrangian submanifold in a homology class $\Gamma$ must break up into components in different homology classes that sum to the original $\Gamma$. This decomposition takes place at a {\em wall of marginal stability} which, for homology classes $\Gamma_1$ and $\Gamma_2$, is defined as the region in moduli space where\footnote{$K$ is the K{\"a}hler potential on the complex structure moduli space, and is defined as $e^{-K}=-i\int_X \Omega\wedge\overline{\Omega} = -i\Pi^\dagger\Sigma_4\Pi$.
}
\begin{equation}
\text{arg}\left(e^{\frac{K}{2}}\int_{\Gamma_1} \Omega(\varphi)\right)=\text{arg}\left(e^{\frac{K}{2}}\int_{\Gamma_2} \Omega(\varphi)\right).
\end{equation}
The (dis)-assembling of special Lagrangian submanifolds across such walls may be organised with split attractor flows.\cite{Denef:2001ix} An investigation of how the networks defined in this paper vary across walls of marginal stability would be very interesting and we hope to return to this problem in the future. For now, we plot a simple example of two constant phase contours, corresponding to different homology classes, that approach their common wall of marginal stability.

\begin{figure}
    \begin{tikzpicture}
    \node[anchor=south west,inner sep=0] (image) at (0,0) {\frame{\includegraphics[width=\textwidth]{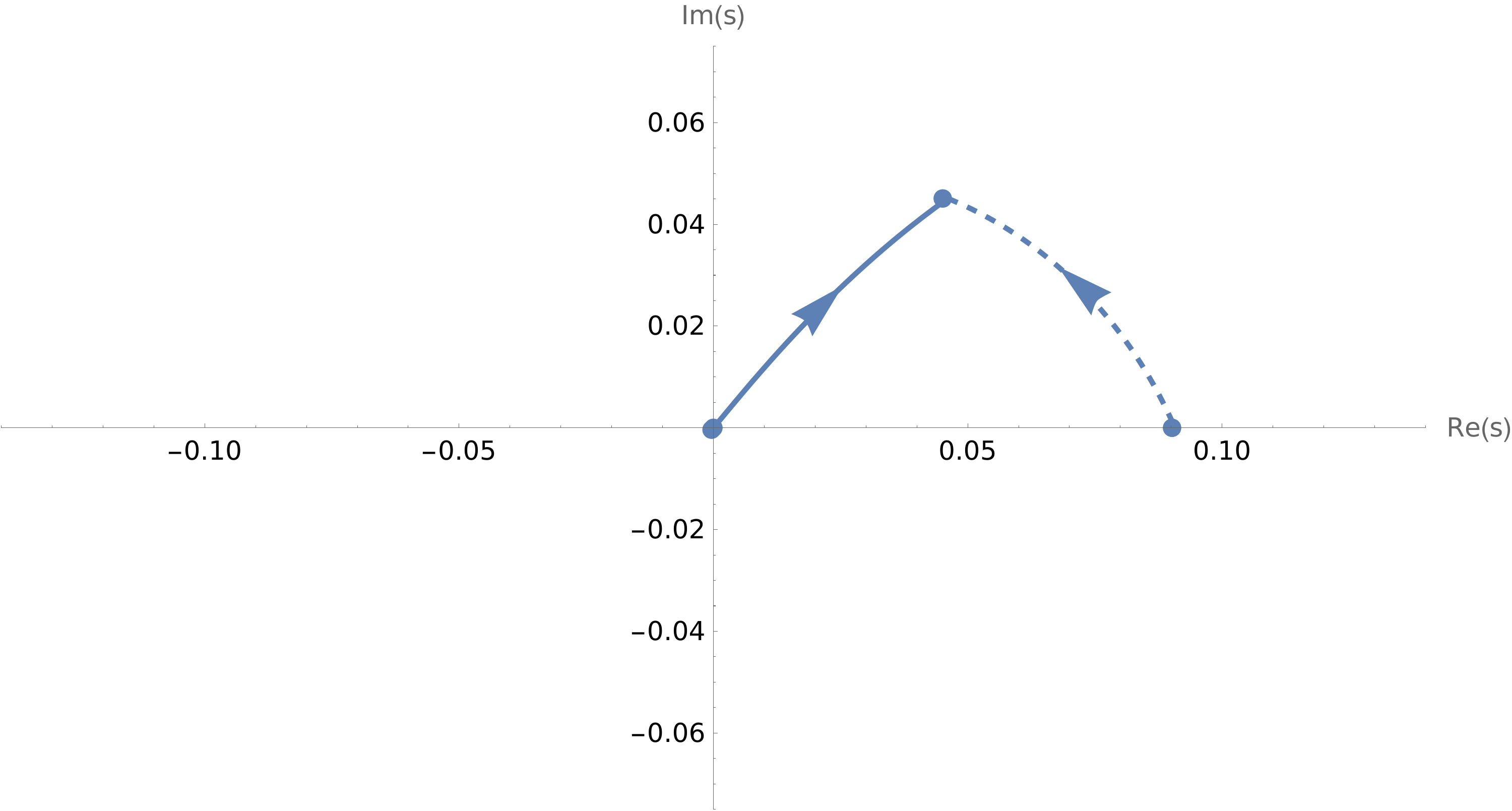}}};
    \begin{scope}[
        x={(image.south east)},
        y={(image.north west)}
    ]
        %\node [white, font=\bfseries] at (0.25,0.65) {$P(X|Y)$};
        \node at (0.89,0.92) {{\large $\varphi=\frac{1}{2}(1+i)s_+^2$}};
        \node at (0.42,0.515) {$s=\frac{\varphi}{s_-}$};
        \node at (0.51,0.44) {$s=0$};
        \node at (0.785,0.4) {$s=s_+$};
        \node at (0.635,0.81) {$ s=\frac{\varphi}{s_+}$};
    \end{scope}
    \end{tikzpicture}

    \vspace{3pt}
        
    \begin{tikzpicture}
    \node[anchor=south west,inner sep=0] (image) at (0,0) {\frame{\includegraphics[width=\textwidth]{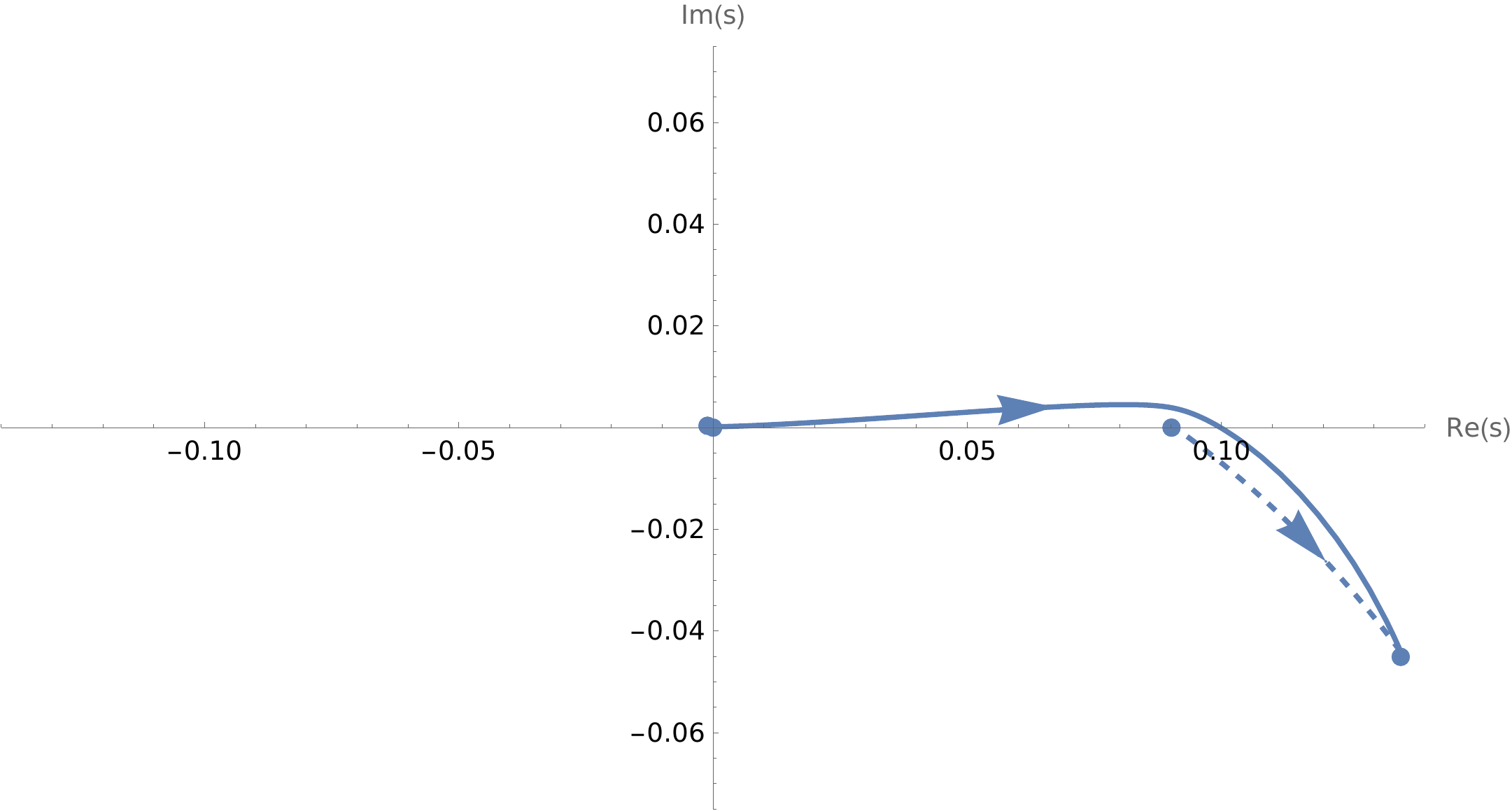}}};
    \begin{scope}[
        x={(image.south east)},
        y={(image.north west)}
    ]
        %\node [white, font=\bfseries] at (0.25,0.65) {$P(X|Y)$};
        \node at (0.89,0.92) {{\large $\varphi=\frac{1}{2}(3-i)s_+^2$}};
        \node at (0.42,0.515) {$s=\frac{\varphi}{s_-}$};
        \node at (0.51,0.44) {$s=0$};
        \node at (0.785,0.54) {$s=s_+$};
        \node at (0.94,0.13) {$ s=\frac{\varphi}{s_+}$};
    \end{scope}
    \end{tikzpicture}
    \captionsetup{format=hang}
    \caption{Two plots of constant phase contours at $\varphi=\frac{1}{2}(1+i)s_+^2$ (top plot) and $\varphi=\frac{1}{2}(3-i)s_+^2$ (bottom plot). In both plots, the solid contour from $s=0$ to $s=\frac{\varphi}{s_+}$ has the associated fibre class $(0,1)^T\otimes(1,0)^T$ and it lifts to the homology class $\Gamma_1=(-1,0,0,0)^T$ when considered together with a constant phase contour from $s=\frac{\varphi}{s_-}$ to $s=0$ with associated fibre class $(0,1)^T\otimes (-4,10)^T$ (this constant phase contour is too small to be seen on these plots). In both plots, the dashed contour from $s=s_+$ to $s=\frac{\varphi}{s_+}$ has the associated fibre class $(-1,0)^T\otimes (1,0)$ and lifts to the homology class $\Gamma_2=(0,0,1,0)^T$. At $\varphi=\frac{1}{2}(1+i)s_+^2$, we find that $\text{arg}\left(\int_{\Gamma_1} \Omega(\varphi)\right)\approx 1.61961$ and $\text{arg}\left(\int_{\Gamma_2} \Omega(\varphi)\right)\approx -1.36498$. Similarly, at $\varphi=\frac{1}{2}(3-i)s_+^2$, we find that $\text{arg}\left(\int_{\Gamma_1} \Omega(\varphi)\right) \approx 1.71821$ and $\text{arg}\left(\int_{\Gamma_2} \Omega(\varphi)\right) \approx 2.62224$. This determines the associated phase $\theta$ in \eqref{eq: constant phase condition} for all of these contours.
    }
    \label{fig: anti D0 and D6 contours approaching MS wall}
\end{figure}
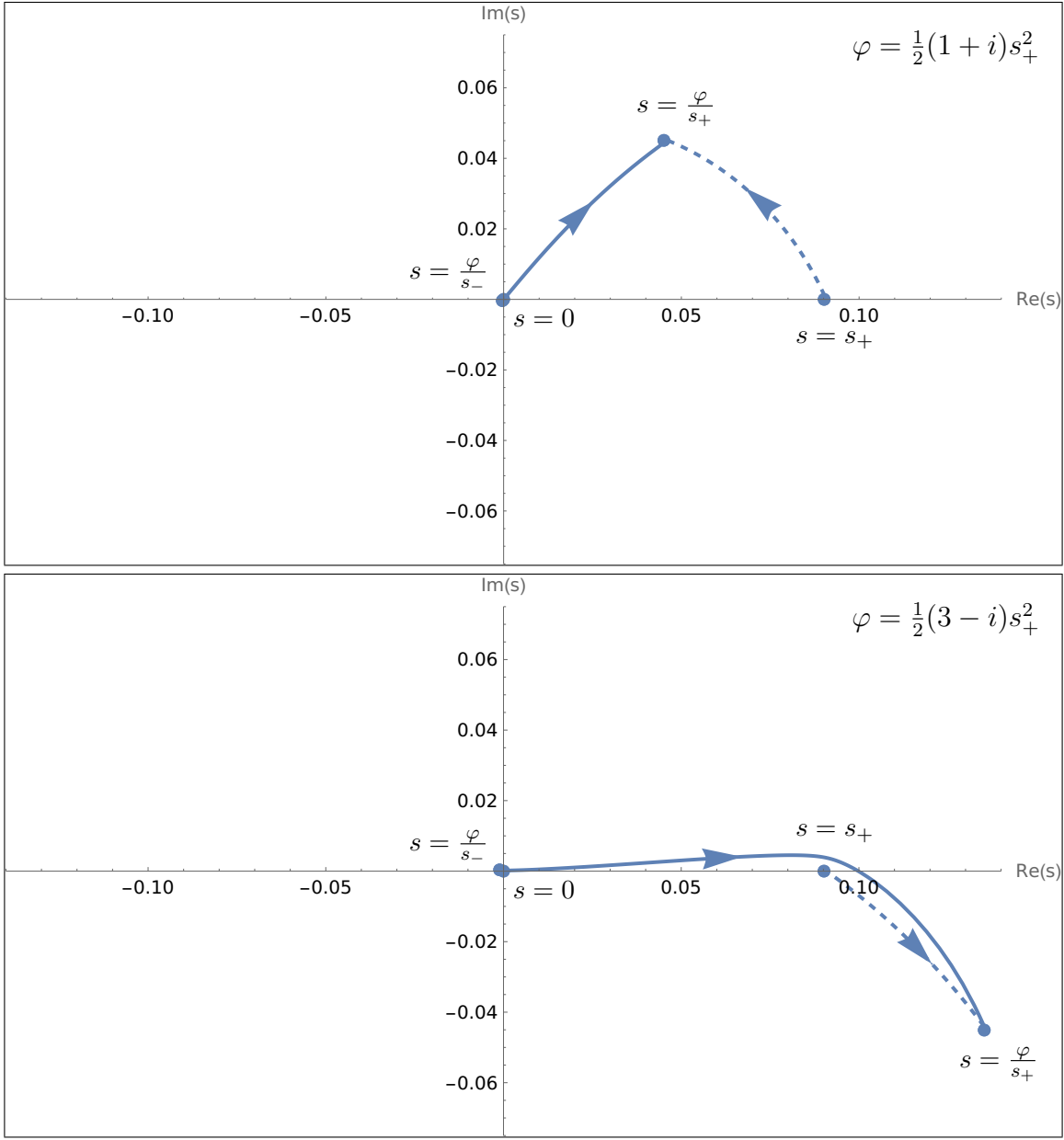

The plots in Figure~\ref{fig: anti D0 and D6 contours approaching MS wall} depict constant phase networks in the homology classes $\Gamma_1=(-1,0,0,0)^T$ (solid contour) and $\Gamma_2=(0,0,1,0)^T$ (dashed contour). Their homology classes are determined by the identities
% \begin{equation}
%     \Gamma_1^T\Sigma_4\Pi(\varphi)=\int_{\frac{\varphi}{s_-}}^0\omega_{\begin{psmallmatrix} 0 \\ 1\\ \end{psmallmatrix}}(s)
%     \omega_{\begin{psmallmatrix} -4 \\ 10\\ \end{psmallmatrix}}\left(\frac{\varphi}{s}\right)
%     \frac{ds}{s}
%     +\int_{0}^{\frac{\varphi}{s_+}}\omega_{\begin{psmallmatrix} 0 \\ 1\\ \end{psmallmatrix}}(s)
%     \omega_{\begin{psmallmatrix} 1 \\ 0\\ \end{psmallmatrix}}\left(\frac{\varphi}{s}\right)
%     \frac{ds}{s}
% \end{equation}
% and
% \begin{equation}
%     \Gamma_2^T\Sigma_4\Pi(\varphi)=\int_{\frac{\varphi}{s_-}}^0\omega_{\begin{psmallmatrix} -1 \\ 0\\ \end{psmallmatrix}}(s)
%     \omega_{\begin{psmallmatrix} 1 \\ 0\\ \end{psmallmatrix}}\left(\frac{\varphi}{s}\right)
%     \frac{ds}{s}
% \end{equation}
\begin{align}
    \begin{split}
        \Gamma_1^T\Sigma_4\Pi(\varphi)&=\int_{\frac{\varphi}{s_-}}^0\omega_{\begin{psmallmatrix} 0 \\ 1\\ \end{psmallmatrix}}(s)
    \omega_{\begin{psmallmatrix} -4 \\ 10\\ \end{psmallmatrix}}\left(\frac{\varphi}{s}\right)
    \frac{ds}{s}
    +\int_{0}^{\frac{\varphi}{s_+}}\omega_{\begin{psmallmatrix} 0 \\ 1\\ \end{psmallmatrix}}(s)
    \omega_{\begin{psmallmatrix} 1 \\ 0\\ \end{psmallmatrix}}\left(\frac{\varphi}{s}\right)
    \frac{ds}{s}\\
    \Gamma_2^T\Sigma_4\Pi(\varphi)&=\int_{s_+}^{\frac{\varphi}{s_+}}\omega_{\begin{psmallmatrix} -1 \\ 0\\ \end{psmallmatrix}}(s)
    \omega_{\begin{psmallmatrix} 1 \\ 0\\ \end{psmallmatrix}}\left(\frac{\varphi}{s}\right)
    \frac{ds}{s}~,
    \end{split}
\end{align}
which we confirm numerically. We see in Figure~\ref{fig: anti D0 and D6 contours approaching MS wall} that the networks in homology classes $\Gamma_1$ and $\Gamma_2$ begin to approach each other as we vary $\varphi$. At the $(\Gamma_1,\Gamma_2)$-wall of marginal stability, two contours from $s_+$ to $\frac{\varphi}{s_+}$ will overlap and we may define a new network by replacing their associated fibre classes with their sum i.e.
\begin{equation}
    \begin{pmatrix} 0 \\ 1\end{pmatrix}\otimes \begin{pmatrix} 1 \\ 0\end{pmatrix}
    +\begin{pmatrix} -1 \\ 0\end{pmatrix}\otimes \begin{pmatrix} 1 \\ 0\end{pmatrix}
    =\begin{pmatrix} -1 \\ 1\end{pmatrix}\otimes \begin{pmatrix} 1 \\ 0 \end{pmatrix}.
\end{equation}
We can do this because $(1,0)^T\otimes\gamma_2$ is a vanishing cycle at the singularity $s_+$ for any choice of $\gamma_2$. Thus, at the $(\Gamma_1,\Gamma_2)$-wall of marginal stability, we find a new network associated with the three-term identity
\begin{align}
    \begin{split}
        (\Gamma_1+\Gamma_2)^T\Sigma_4\Pi(\varphi)=&\int_{\frac{\varphi}{s_-}}^0\omega_{\begin{psmallmatrix} 0 \\ 1\\ \end{psmallmatrix}}(s)
        \omega_{\begin{psmallmatrix} -4 \\ 10\\ \end{psmallmatrix}}\left(\frac{\varphi}{s}\right)
        \frac{ds}{s}
        +\int_{0}^{s_+}\omega_{\begin{psmallmatrix} 0 \\ 1\\ \end{psmallmatrix}}(s)
        \omega_{\begin{psmallmatrix} 1 \\ 0\\ \end{psmallmatrix}}\left(\frac{\varphi}{s}\right)
        \frac{ds}{s}\\
        &+\int_{s_+}^{\frac{\varphi}{s_+}}\omega_{\begin{psmallmatrix} -1 \\ 1\\ \end{psmallmatrix}}(s)
        \omega_{\begin{psmallmatrix} 1 \\ 0\\ \end{psmallmatrix}}\left(\frac{\varphi}{s}\right)
        \frac{ds}{s}.
    \end{split}
\end{align}

Finally, it should be noted that the constant phase contours introduced in this section and the $3$-cycles of the previous section are well-defined when the homology classes in the fibres are non-seperable. Instead of $\gamma^{(i)}_1\otimes\gamma^{(i)}_2\in H_1(\mathcal{E}_s,\mathbb{Z})\otimes H_1(\mathcal{E}_\frac{\varphi}{s},\mathbb{Z})$, we may choose some generic, non-separable $G^{(i)}\in H_1(\mathcal{E}_s,\mathbb{Z})\otimes H_1\left(\mathcal{E}_\frac{\varphi}{s},\mathbb{Z}\right)$ that cannot be written as a tensor product of two homology classes. We may then ask if the union of the $G^{(i)}$ over some contours $\ell^{(i)}$ can be identified with a homology class on a Calabi-Yau threefold. As always, we abuse notation and identify $G^{(i)}$ with a vector of its components in some suitable basis. We then check if an expression of the form
\begin{equation}
    \sum_{i=1}^{N}(G^{(i)})^T\Sigma_2\otimes\Sigma_2\int_{\ell^{(i)}}\omega(s)\otimes\omega\left(\frac{\varphi}{s}\right)\frac{ds}{s}
\end{equation}
is annihilated by the Hadamard product. The constant phase condition \eqref{eq: constant phase condition} is also straightforward to generalise for non-separable fibres, and is given by
\begin{equation}
    G^T\left(\Sigma_2\otimes\Sigma_2\right)\left(\omega(s)\otimes\omega\left(\frac{\varphi}{s}\right)\right)\frac{1}{s}\frac{ds}{dl}=e^{i\theta}~.
\end{equation}
However, if we want to identify the solutions with actual submanifolds, it is important that the homology classes in the fibres are separable.

\subsection{Counting networks?}
We now come to the main unresolved question of this paper. We ask - is it possible to count the number (of families) of solutions of the constant phase condition \eqref{eq: constant phase condition} that lift to a given (torsion free) homology class $\Gamma\in H_3(X,\mathbb{Z})$ at a given $\varphi$? For example, a family of solutions like the one in Figure~\ref{fig: T3 contours at large complex structure} should be counted once. The following heuristic argument suggests that this is possible. Choose some $\Gamma$ as above and a non-singular value of $\varphi$. This determines a phase $\theta=\text{arg}\left(\int_\Gamma \Omega(\varphi)\right)$ that we compute with the fourth order Picard-Fuchs equation (the Hadamard product). We assume that, when a solution of the constant phase condition \eqref{eq: constant phase condition} belongs to a family, we may always choose a contour in this family that starts and ends at singularities on the base $\mathbb{CP}^1$, so we restrict our search to such contours. We analytically continue the elliptic periods to one of the singularities and try solving the constant phase condition \eqref{eq: constant phase condition} for $\gamma_1\otimes\gamma_2$ with small integer coefficients. We may compensate for different choices of analytic continuation by choosing $\gamma_1\otimes\gamma_2$ that are related by known monodromy matrices, so we make an arbitrary choice. Note that a choice of $\gamma_1\otimes\gamma_2$ is a choice of four integers and, if those integers are too large, the factor
\begin{equation}
    \omega_{\gamma_1}(s)\omega_{\gamma_2}\left(\frac{\varphi}{s}\right)\frac{1}{s}
\end{equation}
will also be large. This means that the absolute value of $\frac{ds}{dl}$ will be correspondingly small and the constant phase contour will be very slow to move away from the starting singularity. Since we want a network of constant phase contours that lift to a given $\Gamma$ at a fixed $\varphi$, the parameter $l$ is bounded by
\begin{equation}
    \label{eq: bound on l for constant phase contours}
    l\leq \left|\int_\Gamma \Omega(\varphi)\right|.
\end{equation}
It follows that, if the components of $\gamma_1\otimes\gamma_2$ are too large, it will be impossible for the constant phase contour to reach another singularity or form a non-trivial homotopy class in ``time" $l$ that satisfies \eqref{eq: bound on l for constant phase contours}.

In principle, one could fix $\Gamma$ and $\varphi$ and then simply perform a brute force computer search for solutions of $\eqref{eq: constant phase condition}$ that begin at a given singularity. One would start at some singularity and iterate over $\gamma_1\otimes\gamma_2$ with increasingly large coefficients until it becomes clear that the solutions will not reach another singularity while satisfying \eqref{eq: bound on l for constant phase contours}. One would then perform this search for all singularities and then, armed with finitely many contours, find all combinations of these contours that lift to the given homology class $\Gamma$. Moreover, since the BPS networks introduced in this section capture information about the topology of the associated special Lagrangian submanifold, via their deformations, one would ultimately hope to be able to count $A$-branes and make contact with known enumerative invariants of the associated compact Calabi-Yau threefold (and its mirror). An analogous problem for exponential networks has been studied in \cite{Banerjee:2022oed}.

\acknowledgments

The author benefited from many helpful conversations with Kilian B\"onisch, Philip Candelas, Xenia de la Ossa, Vasily Golyshev, Amir-Kian Kashani-Poor, Matthew Kerr, Albrecht Klemm, Gregory Moore, Fernando Rodriguez Villegas and Mauricio Romo. The author is also grateful for hospitality and support from the Bethe Centre for Theoretical Physics (Bonn) and The Abdus Salam International Centre for Theoretical Physics (Trieste). The author is primarily supported by the US Department of Energy under grant DE-SC0010008.

 \appendix

\section{$\varphi$-monodromy as a braid}
\label{sec: varphi-monodromy as a braid}

When studying identities of the form
\begin{equation}
\Pi_\Gamma(\varphi)=\sum_{i=1}^{N}\int_{\ell^{(i)}}\omega_{\gamma^{(i)}_1}(s)\omega_{\gamma^{(i)}_2}\left(\frac{\varphi}{s}\right)\frac{ds}{s},
\end{equation}
one must worry about monodromy as $\varphi$ is varied around some singular points. The left hand side simply transforms as $\Pi(\varphi)\mapsto M\Pi(\varphi)$ for some monodromy matrix $M$.\footnote{Which implies that $\Pi_\Gamma(\varphi)\mapsto \Pi_{{M^{-1}}\Gamma}(\varphi)$, because $\Sigma_4M=(M^{-1})^T\Sigma_4$ for $M\in Sp(4,\mathbb{Z})$.}  However, the analytic continuation of the integrals on the right hand side is more subtle. As the parameter $\varphi$ is varied, some of the singularities in the $s$-plane will move around and the contours $\ell^{(i)}$ must be deformed to avoid these singularities. We will focus on the case of closed contours, because any open contour is homotopy equivalent to the composition of a small set of open contours and an arbitrarily complicated closed contour. For example, suppose that $\varphi$ is real and all the singularities in the $s$-plane lie on the real line. An open contour between singularities $s_1$ and $s_2$ is homotopy equivalent to the composition of a path from $s_1$ to a base point $s_0$ in the upper half $s$-plane, a loop based at $s_0$ and a path from $s_0$ to $s_2$ along the upper half plane. The action of $\varphi$-monodromy on an open contour from the base point $s_0$ to a singularity on the real line is readily computed by hand, and there are finitely many such contours to consider. However, the closed loop based at $s_0$ can be arbitrarily complicated. 

\begin{figure}[h]
    % [trim={left bottom right top},clip]
    \begin{tikzpicture}
    \node[anchor=south west,inner sep=0] (image) at (0,0) {\frame{\includegraphics[width=\textwidth,trim={0 19cm 0 1cm},clip]{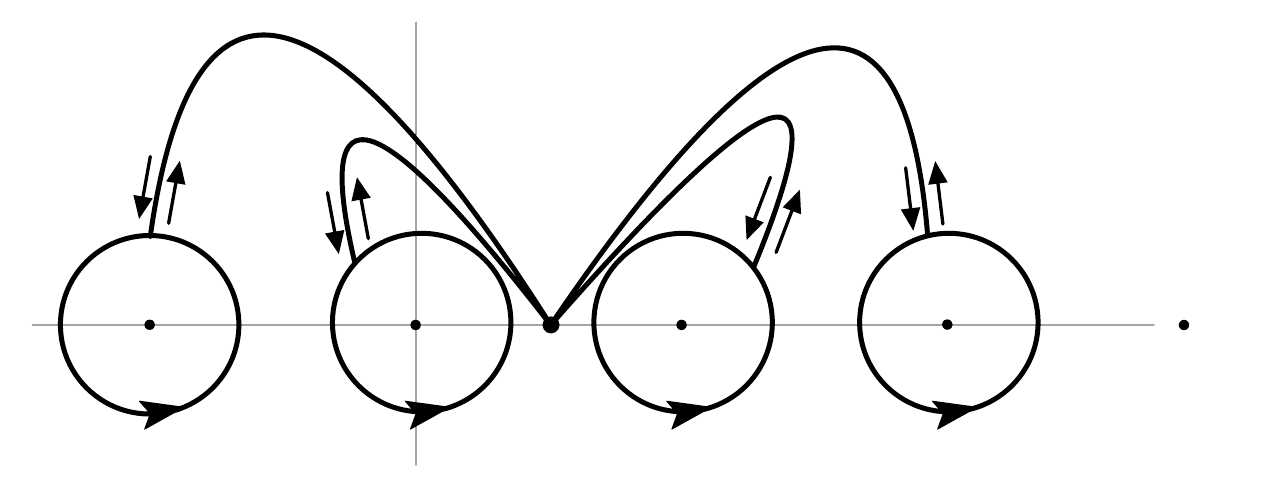}}};
    \begin{scope}[
        x={(image.south east)},
        y={(image.north west)}
    ]
        %\node [white, font=\bfseries] at (0.25,0.65) {$P(X|Y)$};
        \node at (0.126,0.27) {$\varphi=-1$};
        \node[fill=white] at (0.324,0.27) {$\varphi=0$};
        \node at (0.541,0.27) {$\varphi=s_+^2$};
        \node at (0.749,0.27) {$\varphi=s_-^2$};
        \node at (0.935,0.27) {$\varphi=\infty$};
        \node at (0.12,0.82) {$L_{-1}$};
        \node at (0.25,0.70) {$L_{0}$};
        \node at (0.645,0.70) {$L_{s_+^2}$};
        \node at (0.73,0.82) {$L_{s_-^2}$};
        \node at (0.436,0.27) {$\varphi_0$};
    \end{scope}
    \end{tikzpicture}
    \captionsetup{format=hang}
    \caption{Generators of $\pi_1\left(\mathbb{CP}^1\backslash\{-1,0,s_+^2,s_-^2,\infty\},\varphi_0\right)$ in the complex $\varphi$-plane. Note that we have chosen a real basepoint $\varphi_0$ with $0<\varphi_0<s_+^2$.}
    \label{fig: generators of fundamental group for AESZ101}
\end{figure}
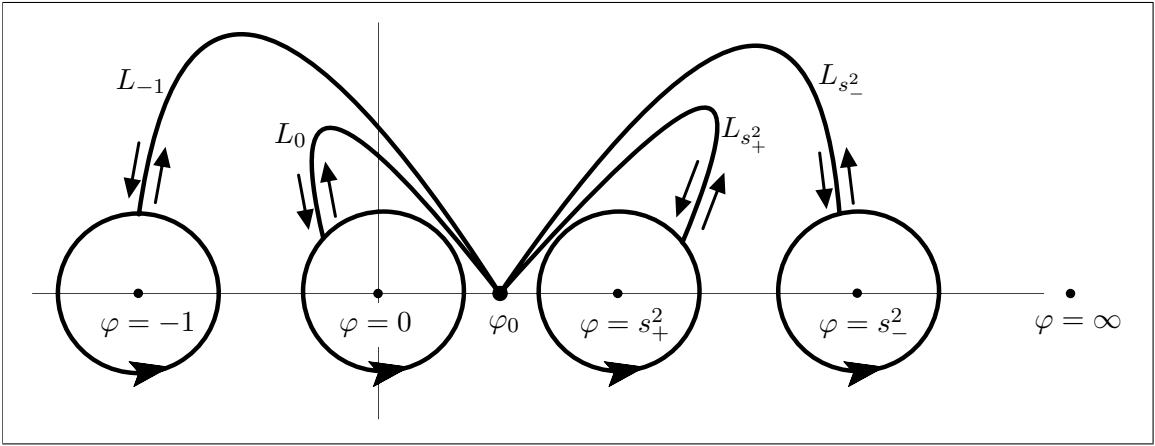

Fortunately, it is not too difficult to quickly compute the action of $\varphi$-monodromy on $\pi_1\left(\mathbb{CP}^1\backslash\{s_-,\frac{\varphi}{s_-},0,\frac{\varphi}{s_+},s_+\},s_0\right)$. To be concrete, consider AESZ 101 and suppose that $\varphi_0$ is real and $0\leq \varphi_0\leq s_+^2$. The choice $\varphi=\varphi_0$ sets all the singularities in the $s$-plane to real values and fixes their ordering from left to right in the $s$-plane. We fix a set of generators of $\pi_1\left(\mathbb{CP}^1\backslash\{-1,0,s_+^2,s_-^2,\infty\},\varphi_0\right)$ as in Figure~\ref{fig: generators of fundamental group for AESZ101}.

By drawing contours on the complex $s$-plane, one might see that the homotopy class $\ell_{\frac{\varphi}{s_+}}\in \pi_1\left(\mathbb{CP}^1\backslash\{s_-,\frac{\varphi}{s_-},0,\frac{\varphi}{s_+},s_+\},s_0\right)$ is mapped to $\ell_{\frac{\varphi}{s_-}}\ell_0\ell_{\frac{\varphi}{s_+}}\ell_0^{-1}\ell_{\frac{\varphi}{s_-}}^{-1}$ when $\varphi$ is transported along $L_0$ (see Figure~\ref{fig: generators of fundamental group of s-plane for AESZ101} for the generators $\ell_{s_*}$).  However, this quickly becomes impractical. A more systematic approach is to identify $L_0$ with a braid that acts on $\pi_1\left(\mathbb{CP}^1\backslash\{s_-,\frac{\varphi}{s_-},0,\frac{\varphi}{s_+},s_+\},s_0\right)$.\cite{artin1947theory}

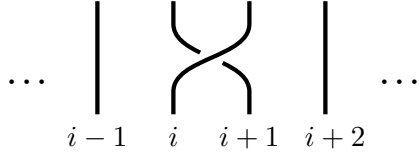
\begin{figure}
\begin{center}
\begin{tikzpicture}
\pic[
braid/.cd,
number of strands=4,
ultra thick,
%thick,
%strand 1/.style={red},
%strand 2/.style={green},
%strand 3/.style={blue},
gap=0.1,
% control factor=0,
% nudge factor=0,
name prefix=braid,
] {braid = {s_2^{-1} }};
\node[at=(braid-1-e),below=1pt] {$i-1$};
\node[at=(braid-2-e),below=1pt] {$i+1$};
\node[at=(braid-3-e),below=1pt] {$i$};
\node[at=(braid-4-e),below=1pt] {$~~i+2$};
\node[at=(braid-2-1),above=5pt,left=70pt] {$\boldsymbol{\cdots}$};
\node[at=(braid-2-1),above=5pt,right=45pt] {$\boldsymbol{\cdots}$};
\end{tikzpicture}
\end{center}
\captionsetup{format=hang}
\caption{The braid represented by the generator $\sigma_i\in B_n$.}
\label{fig: the braid represented by sigma_i}
\end{figure}

\begin{figure}
\begin{center}
\begin{tikzpicture}
\pic[
braid/.cd,
number of strands=4,
ultra thick,
%thick,
%strand 1/.style={red},
%strand 2/.style={green},
%strand 3/.style={blue},
gap=0.1,
% control factor=0,
% nudge factor=0,
name prefix=braid,
] {braid = {s_3^{-1}s_2^{-1} }};
\node[at=(braid-1-e),below=1pt] {$i-1$};
\node[at=(braid-2-e),below=1pt] {$i+1$};
\node[at=(braid-3-e),below=1pt] {$~~i+2$};
\node[at=(braid-4-e),below=1pt] {$i$};
\node[at=(braid-2-2),above=5pt,left=70pt] {$\boldsymbol{\cdots}$};
\node[at=(braid-2-2),above=5pt,right=45pt] {$\boldsymbol{\cdots}$};
\end{tikzpicture}
\end{center}
\captionsetup{format=hang}
\caption{The braid represented by $\sigma_i\sigma_{i+1}\in B_n$.}
\label{fig: the braid represented by sigma_isigma_j}
\end{figure}
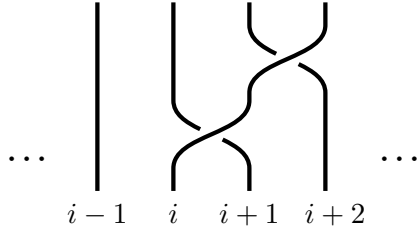

The $n^{\text{th}}$ braid group $B_n$ may be defined in terms of generators $\sigma_i$ as
\begin{equation}
    B_n=\Bigg\langle \sigma_1,\sigma_2,\ldots,\sigma_{n-1}~
    \begin{array}{|r l l}
        \sigma_i\sigma_j&=\sigma_j\sigma_i &~\text{if}~2\leq|i-j|\\
        \sigma_i\sigma_{i+1}\sigma_i&=\sigma_{i+1}\sigma_i\sigma_{i+1} &~\text{if}~1\leq i \leq n-2
    \end{array}
    \Bigg\rangle,
\end{equation}
where each generator $\sigma_i$ should be understood as the braid with the $i^{\text{th}}$ strand moved over the $(i+1)^{\text{th}}$ strand and all other strands left unchanged. This is represented by the braid diagram in Figure~\ref{fig: the braid represented by sigma_i}.
Our conventions are such that time increases as one moves upwards along a braid i.e. the braid $\sigma_i\sigma_{i+1}$ is represented by the braid in Figure~\ref{fig: the braid represented by sigma_isigma_j}.
Label the generators of the free group $F_n$ as $x_i$, where $i\in\{1,2,\ldots,n\}$. The braid group $B_n$ acts on $F_n$ from the right. In terms of generators, this action is defined as
\begin{equation}
    \label{eq: action of braids on fundamental group}
    x_j\sigma_i=\begin{cases}
        x_jx_{j+1}x_j^{-1} &\text{if}~i=j\\
        x_{j-1}            &\text{if}~i=j-1\\
        x_j                &\text{otherwise}.
    \end{cases}
\end{equation}
One should understand this group action as follows. A braid intersects the complex plane $\mathbb{C}$ transversely. At the start of the braid, loops around the points of intersection define an element of the fundamental group of the $n$-times punctured plane for some choice of base point. This is isomorphic to the free group $F_n$. Now one may transport such a loop upwards along a braid and express the transported loop in terms of the original generators of $F_n$. In this way, every braid defines an automorphism of $F_n$.

\begin{figure}[h]
\begin{center}
\begin{tikzpicture}
\pic[
braid/.cd,
number of strands=5,
ultra thick,
gap=0.1,
% control factor=0,
% nudge factor=0,
name prefix=braid,
] {braid = { s_2^{-1} s_3^{-1} s_2^{-1} s_2^{-1} s_3^{-1} s_2^{-1} }};
\node[at=(braid-1-e),below=1pt] {$s_-$};
\node[at=(braid-2-e),below=1pt] {$\frac{\varphi}{s_-}$};
\node[at=(braid-3-e),below=1pt] {$0$};
\node[at=(braid-4-e),below=1pt] {$\frac{\varphi}{s_+}$};
\node[at=(braid-5-e),below=1pt] {$s_+$};
\node[at=(braid-1-s),above=1pt] {$s_-$};
\node[at=(braid-2-s),above=1pt] {$\frac{\varphi}{s_-}$};
\node[at=(braid-3-s),above=1pt] {$0$};
\node[at=(braid-4-s),above=1pt] {$\frac{\varphi}{s_+}$};
\node[at=(braid-5-s),above=1pt] {$s_+$};
\end{tikzpicture}
\end{center}
\captionsetup{format=hang}
\caption{$\mu(L_0)=(\sigma_2\sigma_3\sigma_2)^2$ as a braid diagram.}
\label{eq: braid associated to phi monodromy of AESZ 101 around 0}
\end{figure}
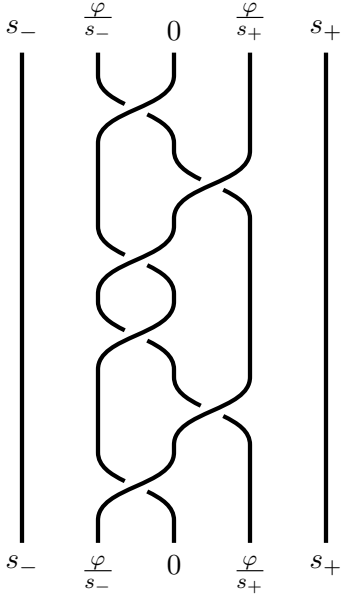

We compute the homomorphism $\mu:\pi_1\left(\mathbb{CP}^1\backslash\{-1,0,s_+^2,s_-^2,\infty\},\varphi_0\right)\rightarrow B_5$ by examining the way singularities in the $s$-plane move (see Figure~\ref{eq: braid associated to phi monodromy of AESZ 101 around 0} for an example). We find that $\mu$ maps the generators in Figure~\ref{fig: generators of fundamental group for AESZ101} to the braids 
\begin{equation}
    \begin{alignedat}{2}
           &\mu(L_{-1})&&=(\sigma_2\sigma_3\sigma_2)(\sigma_4^2\sigma_1^2)(\sigma_2\sigma_3\sigma_2)^{-1}\\
        &\mu(L_0)&&=(\sigma_2\sigma_3\sigma_2)^2\\
        &\mu(L_{s_+^2})&&=\sigma_4^{2}\\
        &\mu(L_{s_-^2})&&=\sigma_1^2~.
    \end{alignedat}
\end{equation}
Note that the image of $\mu$ lies in the subgroup of pure braids i.e. braids in which the strands start and end in the same order. This, of course, is just a consequence of the fact that we consider closed loops in the $\varphi$-plane, which start and end at the same point.

It is now straightforward to compute the action of $\varphi$-monodromy on an element of $\pi_1\left(\mathbb{CP}^1\backslash\{s_-,\frac{\varphi}{s_-},0,\frac{\varphi}{s_+},s_+,\infty\},s_0\right)$. We label the generators in Figure~\ref{fig: generators of fundamental group of s-plane for AESZ101} from left to right (i.e. $x_1=\ell_{s_-},x_2=\ell_{\frac{\varphi}{s_-}},\ldots$) and use \eqref{eq: action of braids on fundamental group}. This is already implemented in many computer algebra systems e.g. SageMath.\cite{sagemath} As an example, consider the identity \eqref{eq: identity for Hadamard integral associated to vanishing cycle of AESZ 101 at nearest conifold}. Repeated here, it is given by
\begin{equation}
    \left(1,0,0,0\right)\Sigma_4\Pi(\varphi)=\oint_{\ell_{\frac{\varphi}{s_-}} \ell_0 \ell_{\frac{\varphi}{s_+}}}
    \omega_{\begin{psmallmatrix} 0 \\ 1\\ \end{psmallmatrix}}(s)
    \omega_{\begin{psmallmatrix} 0 \\ 1\\ \end{psmallmatrix}}\left(\frac{\varphi}{s}\right)
    \frac{ds}{s}.
\end{equation}
Analytically continuing both sides around $L_{s_-^2}$ leads to the new identity
\begin{equation}
    \left(1,0,0,0\right)\left(M_{s_-^2}^{T}\right)^{-1}\Sigma_4\Pi(\varphi)=\oint_{\left(\ell_{\frac{\varphi}{s_-}} \ell_0 \ell_{\frac{\varphi}{s_+}}\right)\mu\left(L_{s_-^2}\right)}
    \omega_{\begin{psmallmatrix} 0 \\ 1\\ \end{psmallmatrix}}(s)
    \omega_{\begin{psmallmatrix} 0 \\ 1\\ \end{psmallmatrix}}\left(\frac{\varphi}{s}\right)
    \frac{ds}{s},
\end{equation}
where $M_{s_-^2}$ is the monodromy matrix computed in \eqref{eq: monodromy matrices for AESZ 101} and
\begin{equation}
    \left(\ell_{\frac{\varphi}{s_-}} \ell_0 \ell_{\frac{\varphi}{s_+}}\right)\mu\left(L_{s_-^2}\right)=\ell_{s_-}\ell_{\frac{\varphi}{s_-}}\ell_{s_-}^{-1}\ell_0\ell_{\frac{\varphi}{s_+}}.
\end{equation}

% \paragraph{Note added.} This is also a good position for notes added
% after the paper has been written.

% Bibliography

%% [A] Recommended: using JHEP.bst file
\bibliographystyle{JHEP}
\bibliography{biblio.bib}

%% or
%% [B] Manual formatting (see below)
%% (i) We suggest to always provide author, title and journal data or doi:
%% in short all the informations that clearly identify a document.
%% (ii) please avoid comments such as "For a review'', "For some examples",
%% "and references therein" or move them in the text. In general, please leave only references in the bibliography and move all
%% accessory text in footnotes.
%% (iii) Also, please have only one work for each \bibitem.

% \begin{thebibliography}{99}

% \bibitem{a}
% Author,
% \emph{Title},
% \emph{J. Abbrev.} {\bf vol} (year) pg.

% \bibitem{b}
% Author,
% \emph{Title},
% arxiv:1234.5678.

% \bibitem{c}
% Author,
% \emph{Title},
% Publisher (year).

% \end{thebibliography}
\end{document}